\documentclass[longbibliography,aps,prb,twocolumn,secnumarabic,amsmath,amssymb,superscriptaddress]{revtex4-1}

\usepackage[colorlinks,bookmarks=false,citecolor=blue,linkcolor=red,urlcolor=blue]{hyperref}
\usepackage{bm}
\usepackage{epsfig}
\usepackage{epstopdf}
\usepackage{color}
\usepackage{multirow}
\usepackage{amsfonts}
\usepackage{amsmath}
\usepackage{amssymb}
\usepackage{subfigure}

\newcommand{\be}{\begin{equation}}
\newcommand{\ee}{\end{equation}}
\newcommand{\ket}[1]{| #1 \rangle} 
\newcommand{\bra}[1]{\langle #1 |} 

\newcommand*\rfrac[2]{{}^{#1}\!/_{#2}} 

\def\bs{\boldsymbol}
\def\vec{\mathbf}
\def\mc{\mathcal}

\begin{document}

\title{Orbital reconstruction in nonpolar tetravalent transition-metal oxide layers}

\author{Nikolay A.~Bogdanov}
\affiliation{Institute for Theoretical Solid State Physics, IFW Dresden, Helmholtzstr.~20, 01069 Dresden,
 Germany}

\author{Vamshi M.~Katukuri}
\affiliation{Institute for Theoretical Solid State Physics, IFW Dresden, Helmholtzstr.~20, 01069 Dresden,
 Germany}

\author{Judit Romh\'anyi}
\affiliation{Institute for Theoretical Solid State Physics, IFW Dresden, Helmholtzstr.~20, 01069 Dresden,
 Germany}

\author{Viktor Yushankhai}
\affiliation{Joint Institute for Nuclear Research, Joliot-Curie 6, 141980 Dubna, Russia}
\affiliation{Institute for Theoretical Solid State Physics, IFW Dresden, Helmholtzstr.~20, 01069 Dresden,
 Germany}

\author{Vladislav Kataev}
\affiliation{Institute for Solid State Research, IFW Dresden, Helmholtzstr.~20, 01069 Dresden, Germany}

\author{Bernd B\"uchner}
\affiliation{Institute for Solid State Research, IFW Dresden, Helmholtzstr.~20, 01069 Dresden, Germany}
\affiliation{Department of Physics, Technical University Dresden, 01062 Dresden, Germany}

\author{Jeroen van den Brink}
\affiliation{Institute for Theoretical Solid State Physics, IFW Dresden, Helmholtzstr.~20, 01069 Dresden,
 Germany}
\affiliation{Department of Physics, Technical University Dresden, 01062 Dresden, Germany}

\author{Liviu Hozoi}
\affiliation{Institute for Theoretical Solid State Physics, IFW Dresden, Helmholtzstr.~20, 01069 Dresden,
 Germany}

\begin{abstract}
A promising route to tailoring the electronic properties of quantum materials and devices rests on the
idea of orbital engineering in multilayered oxide heterostructures.
Here we show that the interplay of interlayer charge imbalance and ligand distortions provides a knob for
tuning the sequence of electronic levels even in intrinsically stacked oxides.
We resolve in this regard the $d$-level structure of layered Sr$_2$IrO$_4$ by electron spin resonance.
While canonical ligand-field theory predicts $g_{\parallel}$-factors $\!<\!2$ for positive tetragonal
distortions as present in Sr$_2$IrO$_4$, the experiment indicates $g_{\parallel}\!>\!2$.
This implies that the iridium $d$ levels are inverted with respect to their normal ordering.
State-of-the-art electronic-structure calculations confirm the level switching in Sr$_2$IrO$_4$, whereas
we find them in Ba$_2$IrO$_4$ to be instead normally ordered.
Given the nonpolar character of the metal-oxygen layers, our findings highlight the tetravalent
transition-metal 214 oxides as ideal platforms to explore $d$-orbital reconstruction in the context of
oxide electronics.
\end{abstract}

\date{October 27, 2014}

\maketitle

Their unique diversity of transport and magnetic properties endows transition-metal (TM) oxides with a
long-term potential for applications in microelectronics and electrical engineering.
Nowadays the search for new or superior properties goes beyond known bulk phases and includes oxide
interfaces and stacked superlattices \cite{heterostruct_heber_2009,heterostruct_mannhart_2010}.
As compared to the bulk material, at interfaces the modification of the nearby surroundings can
significantly affect the valence electronic structure, in particular, the occupation of the $d$-shell
levels \cite{heterostruct_heber_2009, heterostruct_mannhart_2010, OrbRec_chakhalian_2007,
  OrbRec_benckiser_2011, OrbRec_tokura_2012}.
This is often referred to as orbital reconstruction \cite{OrbRec_chakhalian_2007, OrbRec_benckiser_2011,
  OrbRec_tokura_2012} and brings to the fore the most basic aspect in electronic-structure theory: how
energy levels in quantum matter are formed and populated.

A variety of intrinsically stacked crystalline oxides is presently known.
The high-temperature cuprate superconductors \cite{Leggett06}, for example, fall in this category but
also iridates of the type A$_2$IrO$_4$ (A=Sr$^{2+}$, Ba$^{2+}$) which closely resemble undoped cuprates,
both structurally and magnetically \cite{Sr2IrO4_bjkim_08, Sr2IrO4_bjkim_09, Ir214_jkim_2012,
  Ir214_Kim_fermi_arcs_2014, Ir214_Kim_excitonQP_2014}.
Sr$_2$IrO$_4$ has a rather simple crystalline structure displaying stacked, quasi two-dimensional (2D)
IrO$_2$ and double SrO layers.
We shall demonstrate that in this system the occupation of the valence $d$ electronic levels differs from
what is expected in textbook ligand-field theory due to electrostatics that involves both types of
metal-oxygen sheets.
In particular, we show that, as compared to the isostructural cuprate La$_2$CuO$_4$, a different
distribution of ionic charges between the TM-O$_2$ and A-O layers modify the sequence of energy levels
within the $t_{\mathrm{2g}}$ and $e_{\mathrm{g}}$ manifolds and consequently very fundamental physical properties such as the
magnetic $g$ factors, which determine the relation between the magnetic moment and quantum
  number of a magnetic particle.
Our findings are of direct relevance to the field of stacked oxide heterostructures and provide a
guideline on how low-symmetry crystal fields at $d$-metal sites can be altered and potentially engineered
through the appropriate design of successive ionic layers.

To show this we first use electron spin resonance (ESR) measurements to untangle the $5d$-shell
electronic structure of crystalline Sr$_2$IrO$_4$, in particular, the exact order of the Ir $t_{\mathrm{2g}}$
levels.
The single $s\!=\!1/2$ hole present in these $t_{\mathrm{2g}}$ orbitals carries an angular moment
$l_{\mathrm{eff}}\!=\!1$ and is subject to a large spin-orbit coupling (SOC), which in first
approximation results in an effective Ir$^{4+}$ moment, or pseudospin,
$j_{\mathrm{eff}}\!=\!l_{\mathrm{eff}}\!-\!s\!\approx\!1/2$
\cite{Sr2IrO4_bjkim_08,IrO_kitaev_jackeli_09,book_abragam_bleaney}.
We compare the experimental properties of these pseudospins to the ones we have calculated by {\it ab
  initio} quantum chemistry methods.
This combined approach, explored here on a strongly spin-orbit coupled material for the first time,
provides direct access to the spatial anisotropies of the $g$ factors and further to the detailed
microscopic superexchange interactions.
The ESR measurements and theory are found to agree on a quantitative level and moreover undoubtedly show
that the $d$-level ordering in Sr$_2$IrO$_4$ is inverted with respect to the normal ordering in the
sister iridate Ba$_2$IrO$_4$ or the isostructural 214 cuprate superconductors.
The good agreement between the ESR data and the outcome of the computational methodology we describe and
employ here establishes the latter as a reliable tool for the investigation of nontrivial electronic
structures and magnetic couplings.\\

\begin{figure}[!b]
  \includegraphics[width=0.98\columnwidth]{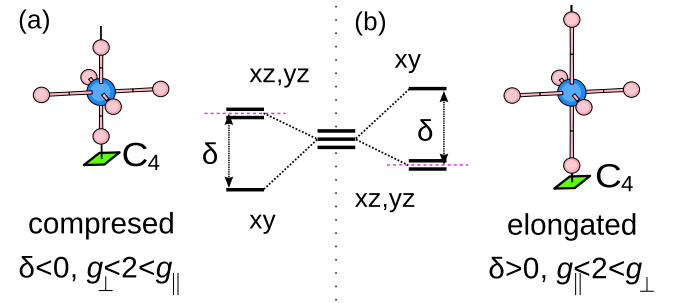}
  \caption{ {\bf TM $\bm{t_{\mathrm{2g}}}$ splittings for tetragonal distortions of the oxygen octahedron sans
      SOC.}
    (a) $z$-axis compression of the octahedron corresponds to a tetragonal splitting $\delta\!<\!0$,
    causes an orbital singlet to be lowest in energy and the $g$ factors to be ordered as
    $g_\perp\!<\!2\!<\!g_\parallel$.
    (b) Elongation of the octahedron ($\delta >0$) causes an orbital doublet to be lowest in energy and
    the $g$ factors to be ordered as $g_\parallel\!<\!2\!<\!g_\perp$.  Purple dashed lines indicate
      the conventional zero level used to define the sign of $\delta$.
  }
    \label{fig:CF}
\end{figure}

\noindent{\bf Results}\\
\noindent{\bf Pseudospins and effective Hamiltonian.} 
Mott-Hubbard physics in $d$-metal compounds has been traditionally associated with
first-series ($3d$) TM oxides.
However, recently, one more ingredient entered the TM-oxide ``Mottness" paradigm -- large
SOC's in $5d$ systems.
SOC in $5d$ and to some extent $4d$ anisotropic oxides modifies the very nature of the
correlation hole of an electron, by admixing the different $t_{\mathrm{2g}}$ components
\cite{IrO_kitaev_jackeli_09,book_abragam_bleaney}, changes the conditions for localization
\cite{Sr2IrO4_bjkim_08}, the criteria of Mottness and further gives rise to new
types of magnetic ground states and excitations \cite{Ir214_jkim_2012,Ir214_Kim_excitonQP_2014}.
While various measurements indicate that indeed spin-orbit-coupled $j_{\mathrm{eff}}\!\approx\!1/2$ states form in
A$_2$IrO$_4$ \cite{Sr2IrO4_bjkim_08, Ir214_jkim_2012, Ba2IrO4_xmcd_2013}, it has been also pointed out
that off-diagonal SOC's may mix into the ground state (GS) wave function substantial amounts of
$t_{\mathrm{2g}}^4e_{\mathrm{g}}^1$ character \cite{SOC_d5_thornley68, SOC_d5_hill71, SOC_d5_andlauer76, SOC_d5_stebler84}.
Such $t_{\mathrm{2g}}^5$--$t_{\mathrm{2g}}^4e_{\mathrm{g}}^1$ many-body interactions were shown to produce remarkable effects in x-ray
absorption and x-ray magnetic circular dichroism (XMCD): the branching ratio between the $L_3$ and $L_2$
Ir $2p$ absorption edges reaches values as large as 4, nearly $50\%$ higher than the 2.75 value for a
``pure" $j_{\mathrm{eff}}\!=\!1/2$ system \cite{Sr2IrO4_xmcd_2012}.
Additionally, low-symmetry noncubic fields produce sizable splittings of the $5d$ $t_{\mathrm{2g}}$ levels, in
some cases close to or even larger than $\sim$1/2 eV \cite{Ir3116_rixs_liu_2012, Ir113_bogdanov_12,
  Ir227_hozoi_14}, and therefore admix the $j_{\mathrm{eff}}\!=\!1/2$ and $j_{\mathrm{eff}}\!=\!3/2$
$t_{\mathrm{2g}}^5$ components \cite{book_abragam_bleaney,SOC_d5_thornley68}.
The structure of the spin-orbit GS depends on both the strength and sign of these splittings.
Interestingly, the best fits of the x-ray absorption and XMCD data are achieved in Sr$_2$IrO$_4$ with a
negative $t_{\mathrm{2g}}$ tetragonal splitting \cite{Sr2IrO4_xmcd_2012}, although the oxygen octahedra in
this material display a distinct  positive tetragonal distortion -- the IrO$_6$ octahedra are
substantially  elongated \cite{Sr214_crawford94} (a negative tetragonal splitting should occur when
the IrO$_6$ octahedra are compressed\cite{book_ballhausen_62,book_abragam_bleaney}, see
Fig.\,\ref{fig:CF}).
This is already a first indication of the level inversion that our ESR measurements and quantum chemistry
calculations show to take place in Sr$_2$IrO$_4$.

The interactions between a pair $\langle ij\rangle$ of nearest-neighbor (NN) 1/2 pseudospins in the
presence of an external magnetic field $\vec{h}$ is given by the effective Hamiltonian
\be\label{SpinHam_2sites}
\mc{H}^{i,j}_{\mathrm{eff}} =
                J \tilde{\vec{S}}_i \cdot \tilde{\vec{S}}_j
                +{\vec D} \cdot {\tilde {\bf S}}_i\times {\tilde {\bf S}}_j
                +\tilde{\vec{S}}_i \cdot \bar{\bar{\bs{\Gamma}}} \cdot \tilde{\vec{S}}_j\,
                +\mu_{\rm B}\!\sum_{k\!=\!i,j} \vec{h} \cdot \bar{\bar{\vec{g}}}_k \cdot \tilde{\vec S}_k,
\ee
where $\tilde{\vec{S}}_i$, $\tilde{\vec{S}}_j$ are pseudospin ($j_{\mathrm{eff}}\!\approx\!1/2$)
operators, $J$ is the isotropic Heisenberg exchange, $\vec{D}\!=\!(0,0,D)$ defines the antisymmetric
Dzyaloshinskii-Moriya (DM) coupling, $\bar{\bar{\bs{\Gamma}}}$ is a symmetric traceless second-rank
tensor describing the symmetric anisotropy and due to the staggered rotation of the IrO$_6$ octahedra the
$\bar{\bar{\vec g}}$ tensor splits for each of the two sites into uniform and staggered components
$\bar{\bar{\vec g}}_k^{\mathrm{u}}\pm\bar{\bar{\vec g}}_k^{\mathrm{s}}$ (see, e.g.,
Refs.\,[\onlinecite{stagg_g_oshikawa_97,SOC_d5_hill71}]).
This effective spin Hamiltonian is of direct relevance to the interpretation of the ESR data.

\begin{figure}[!b]
  \includegraphics[width=0.77\columnwidth]{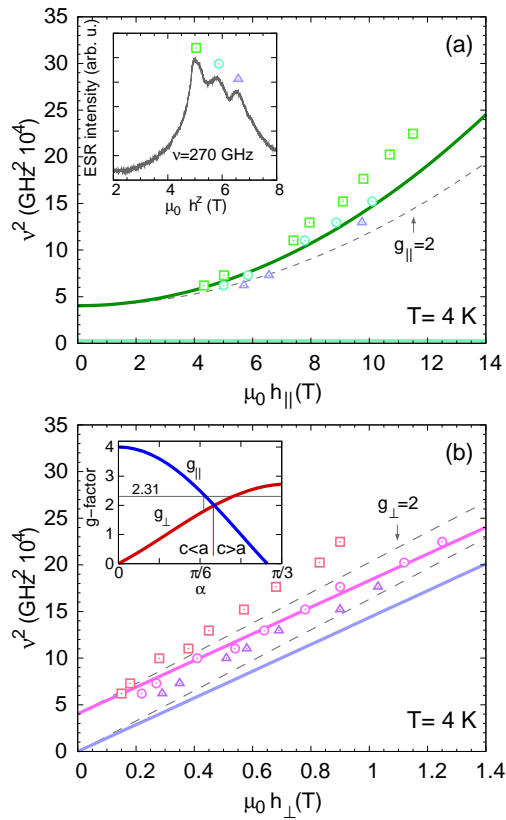}
  \caption{ {\bf ESR data for Sr$\bm{_2}$IrO$\bm{_4}$.}
    (a) Out-of-plane magnetic-field geometries; the insert shows a representative AFR spectrum.
    (b) In-plane magnetic fields; the insert demonstrates the $g$-factor anisotropy as function of the
    tetragonal distortion parameter $\alpha$ (see text).
    Symbols denote experimental data points ---  solid lines are theoretical curves using
    Eqs.\,(\ref{in_pl_ESR},\ref{nu_perp_1},\ref{nu_perp_2}) and the quantum chemically computed $g$
    factors $g_{\parallel}\!=\!2.31$, $g_{\perp}\!=\!1.76$ (see Table~\ref{gfactors_Sr$_2$IrO$_4$_qc});
     dashed lines are calculated assuming the isotropic $g$ factors.
  }
  \label{ESR_v2}
\end{figure}

\noindent{\bf ESR measurements.}
For a single crystal of Sr$_2$IrO$_4$ we observe antiferromagnetic resonance (AFR) modes in the sub-THz
frequency domain \cite{Bahr14} as displayed in Fig.~\ref{ESR_v2}.
There are two modes if $\vec{h}\| z$: a gapless Goldstone mode $\nu_{\| 1}\!=\!0$ and a gapped excitation
\begin{eqnarray}\label{in_pl_ESR}
\nu_{\| 2}^2=\Delta^2+{2 g^2_{\parallel} h^2_\| \tilde J}/({2(J+\Gamma_{\!zz})+\tilde J}),
\end{eqnarray}
where $\Delta=\sqrt{2\tilde J(\tilde J-2(J+\Gamma_{\!zz}))}$, $\tilde J=\sqrt{4D^2+(2 J-\Gamma_{\!zz})^2}$
and $\Gamma_{\!zz}$ couples the ${\tilde S}_{i}^z$ and ${\tilde S}_{j}^z$ components (along the $c$ axis,
perpendicular to the $ab$-plane IrO$_2$ layers \cite{Sr214_crawford94}) in the third term of
Eq.\,(\ref{SpinHam_2sites}) (see Methods for details).
Experimental results are shown for $\nu_{\| 2}$ in Fig.\,\ref{ESR_v2}(a).
The data comprise at $T\!\ll\!T_{\mathrm{N}}\!=\!240$\,K a group of overlapping resonances
[Fig.~\ref{ESR_v2}(a), inset], possibly due to some distribution of internal fields in the sample. Though
revealing some scatter, the $\nu_{\| 2}^2$ AFR data follow approximately a parabolic dependence on $h$
and, most importantly, they lie substantially above the curve corresponding to the free-electron
Land{\'e} factor $g_{\mathrm{e}}\!=\!2$ [dashed line in Fig.~\ref{ESR_v2}(a)].
The experimental dependence $\nu_{\| 2}^2(h^2)$ can be reasonably well modeled with $g_{\parallel}$
values of 2.3--2.45. The solid line in Fig.~\ref{ESR_v2}(a) is obtained by using
$g_{\parallel}\!=\!2.31$, as derived from quantum chemistry calculations that will be discussed later on.
%

Sizable deviations to values larger than 2 of $g_{\parallel}$ is clear indication of the presence of
low-symmetry, noncubic crystal fields.
In the simplest approximation, i.e., restricting ourselves to the Ir$^{4+}$ $t_{\mathrm{2g}}^5$ manifold, the
anisotropic $g$ factors in axial noncubic environment can be expressed up to the sign as
\cite{book_abragam_bleaney}
$
g_{\parallel} = g_c = (2+2k)\cos^2\!\alpha -2\sin^2\!\alpha
$
and 
$
g_{\perp} = g_a = g_b = 2\sin^2\alpha + 2\sqrt{2}k\cos\alpha\sin\alpha, 
$
where $k$ is a covalency reduction factor,
$\alpha\!=\!(1/2)\arctan[2\sqrt{2}\lambda/(\lambda\!-\!2\delta)]$ parameterizes the deviation from
octahedral symmetry, $\lambda$ is the SOC constant and $\delta$ the Ir $t_{\mathrm{2g}}$ splitting.
A plot for the dependence of the diagonal $g$ factors on the distortion parameter $\alpha$ is shown in
the inset to Fig.~\ref{ESR_v2}(b).
For simplicity, $k\!=\!1$ is for the moment assumed but smaller values of $k$ do not bring qualitative
changes.
In cubic symmetry $\delta\!=\!0$, $\alpha_{\mathrm{cub}}\!=\!35.26^\circ$ and the $g$ matrix is isotropic
with $g_{\parallel}\!=\!g_{\perp}\!=\!2$.
According to standard textbooks on ligand-field theory\cite{book_ballhausen_62}, an elongation of the
out-of-plane Ir-O bond induces a positive tetragonal splitting of the Ir $t_{\mathrm{2g}}$ levels, with
$\delta\!>\!0$ and $\alpha\!>\!\alpha_{\mathrm{cub}}$, whereas a bond compression yields $\delta\!<\!0$
and $\alpha\!<\!\alpha_{\mathrm{cub}}$ (see Fig.\,\ref{fig:CF}).
As in Sr$_2$IrO$_4$ the IrO$_6$ octahedra are substantially elongated in the $z$ direction
\cite{Sr214_crawford94}, the $g$ factors are expected to correspond to the case of positive splitting
$\alpha\!>\!\alpha_{\mathrm{cub}}$, see the area to the right of the crossing point shown in the inset to
Fig.~\ref{ESR_v2}(b).
It follows from the plot that $g_{\parallel}\!<\!2$, which obviously contradicts our AFR data for
$\vec{h}\!\parallel\!z$ [Fig.~\ref{ESR_v2}(a)].
The value $g_{\parallel}\!=\!2.31$ used to draw the curve connecting the open circles in
Fig.~\ref{ESR_v2}(a) in fact corresponds to $\alpha\!=\!32.05^{\circ}\!<\!\alpha_{\mathrm{cub}}$ [see the
  inset in Fig.~\ref{ESR_v2}(b)] and indicates that, despite the positive $c$-axis tetragonal distortion,
a counterintuitive  negative tetragonal splitting of the $5d$ $t_{\mathrm{2g}}$ levels is present in
Sr$_2$IrO$_4$.

It should be noted that there is no Goldstone mode for finite in-plane magnetic fields.
The canting angle depends in this case both on the strength of the DM interaction and the applied field.
The two modes are
\be
\label{nu_perp_1}
\nu^2_{\bot 1}=m^2 g^2_{\bot}h^2_\bot+m g_{\bot}h_{\bot}(-4J-\Gamma_{\!zz}+mJ')
\ee
and
\be
\label{nu_perp_2}
\begin{aligned}
\nu^2_{\bot 2}&= m^2 g^2_{\bot}h^2_\bot+m g_{\bot}h_\bot(-4J+5\Gamma_{\!zz}+3mJ')\\
              &+ (-4J+2\Gamma_{\!zz}+2mJ')(3\Gamma_{\!zz}+mJ')\;,
\end{aligned}
\ee 
where $J'\!=\!-4D\sqrt{1-m^2}+2m(2J-\Gamma_{\!zz})$ and $m$ is the in-plane ferromagnetic component of
the effective moments.
To first order in magnetic field, $m$ can be expressed as
\begin{align}
m=-\sqrt{\frac{\tilde J-2J +\Gamma_{\!zz}}{2\tilde J}}- 
 \frac{g_\bot h_\bot}{2\sqrt{6D^2+(2J-\Gamma_{\!zz})^2}}\;,
\label{eq:m}
\end{align}
which holds for weak fields where $m$ remains small.
The first term corresponds to the zero-field canting, arising from the DM interaction, while the second
term shows how this canting evolves with increasing $h_\bot$.
Plots based on Eqs.~(\ref{nu_perp_1},\ref{nu_perp_2}) and the quantum chemically derived interaction
parameters (see below) are displayed in Fig.~\ref{ESR_v2}(b) together with ESR data.

\noindent{\bf Quantum chemistry calculations of $\bm{g}$ factors.}
Results of {\it ab initio} quantum chemistry calculations for the $g$ factors in Sr$_2$IrO$_4$ and in the
structurally related material Ba$_2$IrO$_4$ are listed in Table~\ref{gfactors_Sr$_2$IrO$_4$_qc}.
Our computational scheme follows the prescription of Bolvin \cite{g_bolvin_06} and Vancoillie {\it et
  al.} \cite{g_vancoillie_07}.
It maps the   
matrix elements (ME's) of the {\it ab initio} Zeeman Hamiltonian
%
$
\mc{H}_{\rm Z} = -\bs{\mu}\cdot \vec{h} = \mu_{\rm B}(\vec{L} + g_{\mathrm{e}} \vec{S})\cdot\vec{h}  
$
%
onto the ME's of the effective pseudospin Hamiltonian
%
$
\mc{H}_{\rm S} = \mu_{\rm B}\vec{h} \cdot \bar{\bar{\vec{g}}} \cdot \tilde{\vec S} 
$,
%
where $\bs{\mu}$, $\vec{L}$ and $\vec{S}$ are magnetic moment, angular-momentum and spin operators,
respectively.
The spin-orbit GS wave functions are computed either at the complete-active-space self-consistent-field
(CASSCF) or multireference configuration-interaction (MRCI) level of theory \cite{book_QC_00}, as
described in Ref.\,[\onlinecite{SOC_molpro}] and using the  MOLPRO quantum chemistry package
\cite{molpro_brief}.
All necessary angular-momentum ME's are calculated as well with  MOLPRO (see Methods).
In a first set of calculations, only the three $t_{\mathrm{2g}}$ orbitals at a given Ir site and five electrons
were considered in the active space.
The self-consistent-field optimization was carried out for the corresponding $^2T_{\mathrm{2g}}$($t_{\mathrm{2g}}^5$)
state.
We use here the more convenient notations associated to $O_h$ symmetry, although the calculations were
performed for the actual experimental geometry, with point-group symmetry lower than octahedral.
Inclusion of SOC yields in this case a set of three Kramers doublets (KD's), see Table
\ref{gfactors_Sr$_2$IrO$_4$_qc}.

\begin{table}[!t]
\caption{
{\bf $\bm{g}$ factors for Sr$\bm{_2}$IrO$\bm{_4}$ and Ba$\bm{_2}$IrO$\bm{_4}$.}
Results of many-body quantum chemistry calculations are shown.
The left column displays the electron configurations entering the spin-orbit treatment.
Only the high-spin sextet state is considered out of the $t_{\mathrm{2g}}^{3}e_{\mathrm{g}}^{2}$ manifold.
Final results are indicated in bold face.
}
\label{gfactors_Sr$_2$IrO$_4$_qc}
\begin{ruledtabular}
\begin{tabular}{lcccc}
\multirow{2}{*}{States considered}          &\multicolumn{2}{c}{CASSCF}    &\multicolumn{2}{c}{MRCI}\\
\cline{2-5}
                          &g$_{\perp}$ &g$_{\parallel}$  &g$_{\perp}$   &g$_{\parallel}$ \\[0.03cm]
\hline
\\[-0.35cm]
Sr$_2$IrO$_4$ ($\delta\!=-155$\,meV)\,: & & & \\
~$t_{\mathrm{2g}}^5$ (3 KD's)        &1.67       & 2.25         &1.60         &2.35 \\
~$t_{\mathrm{2g}}^5$, $t_{\mathrm{2g}}^4e_{\mathrm{g}}^1$, $t_{\mathrm{2g}}^3e_{\mathrm{g}}^2$ (27 KD's)
                          &1.81       &2.27          &{\bf 1.76}   &{\bf 2.31} \\
\\[-0.35cm]
Ba$_2$IrO$_4$ ($\delta\!=\!65$\,meV)\,: & & & \\
~$t_{\mathrm{2g}}^5$ (3 KD's)        &2.00       &1.61          &2.01         &1.60 \\
~$t_{\mathrm{2g}}^5$, $t_{\mathrm{2g}}^4e_{\mathrm{g}}^1$, $t_{\mathrm{2g}}^3e_{\mathrm{g}}^2$ (27 KD's)
                          &2.09       &1.77          &{\bf 2.10}   &{\bf 1.76} \\
\end{tabular}  
\end{ruledtabular}
\end{table}

Subsequently we performed calculations with larger active spaces, including also the Ir $e_{\mathrm{g}}$ orbitals.
One $t_{\mathrm{2g}}^5$ ($^2T_{\mathrm{2g}}$) plus four $t_{\mathrm{2g}}^4e_{\mathrm{g}}^1$ ($^2\!A_{\mathrm{2g}}$, $^2T_{\mathrm{1g}}$, $^2\!E_{\mathrm{g}}$, and
$^2T_{\mathrm{2g}}$) spin doublets, two spin quartets [$^4T_{\mathrm{1g}}$($t_{\mathrm{2g}}^4e_{\mathrm{g}}^1$) and
  $^4T_{\mathrm{2g}}$($t_{\mathrm{2g}}^4e_{\mathrm{g}}^1$)] and one spin sextet [$^6\!A_{\mathrm{1g}}$($t_{\mathrm{2g}}^3e_{\mathrm{g}}^2$)] entered here the
spin-orbit treatment.
The orbitals were optimized for an average of all these terms.

The effect of enlarging the active space to include $t_{\mathrm{2g}}^4e_{\mathrm{g}}^1$ terms in the reference wave function
is in the range of $10\%$, in line with earlier semi-empirical estimates for $4d^5$ and $5d^5$ systems
\cite{SOC_d5_thornley68,SOC_d5_hill71,SOC_d5_andlauer76,SOC_d5_stebler84}.
Most importantly, the calculations yield a negative tetragonal splitting of the Ir $t_{\mathrm{2g}}$ levels in
Sr$_2$IrO$_4$, $\delta\!=-155$ meV by MRCI, and positive $t_{\mathrm{2g}}$ splitting in Ba$_2$IrO$_4$ (see
Table~\ref{gfactors_Sr$_2$IrO$_4$_qc} and Methods).
Similar signs, negative in Sr$_2$IrO$_4$ and positive in Ba$_2$IrO$_4$, but much larger magnitudes
($\approx\!0.7$ eV) are found for the computed Ir $e_{\mathrm{g}}$ splittings (not shown in
Table~\ref{gfactors_Sr$_2$IrO$_4$_qc}).

Taken together, the ESR and quantum chemistry results unequivocally point at an anomalous order of the
split Ir 5$d$ levels in Sr$_2$IrO$_4$, related to the important role of the extended crystalline
surroundings in generating low-symmetry fields that compete with ``local'' distortions of the ligand
cage.
Similar effects were found by {\it ab initio} calculations on the 214 layered rhodate Sr$_2$RhO$_4$
\cite{RhIr214_vmk_14} and the 227 pyrochlore iridates \cite{Ir227_hozoi_14}.
In contrast, in Ba$_2$IrO$_4$, the stretch of the apical Ir-O bonds is strong enough \cite{Ba214_okabe11}
to overcome the longer-range electrostatics, turning the tetragonal $t_{\mathrm{2g}}$ positive again, as discussed
in more detail in the following.
Consequently, the structure of the $\bar{\bar{\vec g}}$ tensor in Ba$_2$IrO$_4$ is qualitatively
different, with $g_{\perp}\!>\!2$ and $g_{\parallel}\!<\!2$ (see Table~\ref{gfactors_Sr$_2$IrO$_4$_qc}),
the ordering that one normally expects and encounters for elongated octahedra
\cite{book_abragam_bleaney}.

\noindent{\bf Exchange couplings from quantum chemistry.} 
\begin{figure}[!b]
\includegraphics[width=0.9\columnwidth]{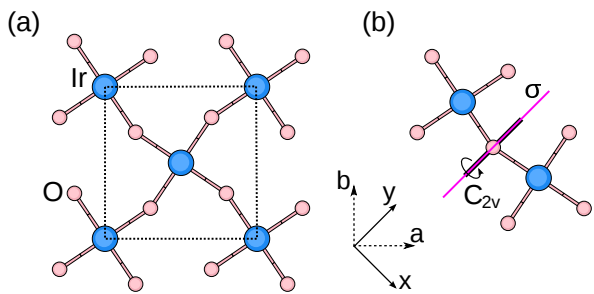}
\caption{
{\bf Planar IrO$\bm{_2}$ network in Sr$\bm{_2}$IrO$\bm{_4}$.}
(a) Coordination of the Ir site.
Dashed lines show the boundaries of the crystallographic unit cell within a given IrO$_2$ layer.
(b) The point-group symmetry of the [Ir$_2$O$_{11}$] block is $C_{\mathrm{2v}}$; associated symmetry elements are
indicated in the figure.
}
\label{fig:symm}
\end{figure}
To obtain {\it ab initio} quantum chemistry values for the inter-site effective magnetic couplings in
Sr$_2$IrO$_4$ [see Eq.\,(\ref{SpinHam_2sites})], we carried out additional calculations on larger
clusters that incorporate two $5d^5$ sites.
The two-octahedra cluster has $C_{\mathrm{2v}}$ symmetry (see Fig.\,\ref{fig:symm}), which implies a diagonal form
for $\bar{\bar{\bf \Gamma}}$ and $\bar{\bar{\vec g}}_i^{\mathrm{u}}$ in  equation (\ref{SpinHam_2sites}) (see
Methods).
By one-to-one correspondence between the ME's of the {\it ab initio} Hamiltonian
\be
\label{QC_Hamm_2sites}
\mc{H}^{i,j} = \mc{H}_{\rm SR}^{i,j} + \mc{H}_{\rm SO}^{i,j} + \mc{H}_{\rm Z}^{i,j}
\ee
and the ME's of the effective spin Hamiltonian (\ref{SpinHam_2sites}) in the basis of the lowest four
spin-orbit states defining the magnetic spectrum of two NN octahedra,
we can derive in addition to the $g$ factors the strengths of the Heisenberg and anisotropic intersite
couplings.
In  equation (\ref{QC_Hamm_2sites}), $\mc{H}_{\rm SR}^{i,j}$ is the scalar-relativistic Born-Oppenheimer
Hamiltonian, $\mc{H}_{\rm SO}^{i,j}$ describes spin-orbit interactions \cite{SOC_molpro} and
$\mc{H}_{\rm Z}^{i,j}$ is the two-site Zeeman Hamiltonian.

Diagonalization of  the spin Hamiltonian (\ref{SpinHam_2sites}) provides the expected singlet
$\ket{\tilde s}$ and three (split) triplet components $\ket{t_{x}}$, $\ket{t_{y}}$ and $\ket{\tilde
  t_{z}}$.
Due to the DM interaction, $\ket{\tilde s}$ and $\ket{\tilde t_{z}}$ are admixtures of ``pure''
$\ket{0,0}$ and $\ket{1,0}$ spin functions.
Our mapping procedure yields $J\!\approx\!48$ meV, somewhat lower than $J$ values of 55--60 meV derived
from experiment \cite{Ir214_jkim_2012,Sr214_johnston94}, and a ratio between the DM and Heisenberg
couplings $D/J\!=\!0.25$, in agreement with estimates based on effective superexchange models
\cite{IrO_kitaev_jackeli_09, Ir214_BHKim_2012, Ir214_perkins_14} and large enough to explain the nearly
rigid rotation of magnetic moments that is observed when the IrO$_6$ octahedra revolve
\cite{Boseggia13,Ye13}.
{\it Ab initio} results for the NN anisotropic couplings $\bar{\bar{\bf \Gamma}}$, also relevant for a
detailed understanding of the magnetic properties of Sr$_2$IrO$_4$, are shown as well in
Table~\ref{2sites_results}.
In our convention the $x$ axis is taken along the Ir-Ir link, i.e., it coincides with the $\langle
110\rangle$ crystallographic direction \cite{Sr214_crawford94}, and $z\!\parallel\!c$.
We obtain $\Gamma_{\!xx}\!\approx\!\Gamma_{\!zz}$, which then allows to recast the Heisenberg and
symmetric anisotropic terms in (\ref{SpinHam_2sites}) as
$(J - \Gamma_{\!yy}/2)\,\tilde{\vec{S}}_i \cdot \tilde{\vec{S}}_j +(3\Gamma_{\!yy}/2)\,\tilde{S}_i^y\,\tilde{S}_j^y$,
with
$\Gamma_{\!xx}\!=\!\Gamma_{\!zz}\!=-\Gamma_{\!yy}/2$.
Equally interesting, for no rotation of the IrO$_6$ octahedra and straight Ir-O-Ir bonds in
Ba$_2$IrO$_4$, it is $\Gamma_{\!yy}$ and $\Gamma_{\!zz}$ which are approximately the same, providing a
realization of the compass-Heisenberg model \cite{Ba214_vmk_14,Ir214_perkins_14} since the DM coupling is
by symmetry 0 in that case.

\begin{table}[!t]
\caption{
{\bf Nearest-neighbonr magnetic couplings in Sr$\bm{_2}$IrO$\bm{_4}$.}
Results of spin-orbit MRCI calculations on two-octahedra clusters are displayed (meV).
$\Gamma_{\!xx}\!+\!\Gamma_{\!yy}\!+\!\Gamma_{\!zz}\!=\!0$ since $\bar{\bar{\bf \Gamma}}$ is traceless.
}
\label{2sites_results}
\begin{ruledtabular}
   \begin{tabular}{ccccc}
   $J$  &$D$       &$\Gamma_{\!xx}$ &$\Gamma_{\!yy}$ &$\Gamma_{\!zz}$ \\
\hline
   &\\[-0.25cm]
   47.8 &$\pm$11.9 &0.42            &--0.84          &0.42 \\
\end{tabular}
\end{ruledtabular}
\end{table}

The two-site magnetic Hamiltonian (\ref{SpinHam_2sites}) features in-plane symmetric-anisotropy couplings
$\Gamma_{\!xx}$ and $\Gamma_{\!yy}$
which were not considered in previous studies \cite{IrO_kitaev_jackeli_09,Bahr14}.
In the presence of two-sublattice order, terms containing these couplings cancel each other in the
mean-field energy but they are in general relevant for pseudospin fluctuations and excitations.
Using spin-wave theory and effective parameters derived from the quantum chemistry calculations, we
nicely reproduce the correct GS and character of the modes, as shown in Fig.~\ref{ESR_v2}.
To reproduce the experimental zero-field gap, in particular, we used $J$, $D$ and $g$-factor values as
listed in Table\,\ref{2sites_results} and a somewhat larger $\Gamma_{\!zz}$ parameter of 0.98 meV.
To leading order, the dependence of $\nu^2_{\bot 1}$ and $\nu^2_{\bot 2}$ on $h$ is linear, see
Eqs.\,(\ref{nu_perp_1},\ref{nu_perp_2}), and the slope is proportional to $m$.
At low fields ($\leq\!1$ T), $m$ can be actually replaced with its field independent value \cite{Bahr14}.
Using the MRCI coupling constants, the first term in  equation (\ref{eq:m}) then yields a moment
$m\!\approx\!0.12\mu_{\mathrm{B}}$, in good agreement with the experiment \cite{Sr2IrO4_bjkim_09,Sr214_cao98}.
\\

\begin{figure}[!b]
\includegraphics[width=0.9\columnwidth]{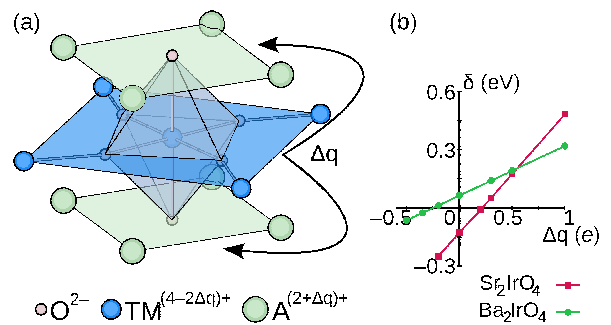}
\caption{
{\bf Effect of interlayer charge imbalance in A$\bm{_2}$IrO$\bm{_4}$ iridates.}
(a) The nearby surroundings of TM sites in A$_2$TMO$_4$ layered perovskites.
In test calculations one can assign the adjacent (in-plane) TM ions the formal charge
$Q_{\mathrm{TM}}\!-\!2\Delta q$, which is compensated by assigning the NN A sites the charge
$Q_{\mathrm{A}}\!+\!\Delta q$.
(b) Tetragonal crystal-field energy splitting between $t_{\mathrm{2g}}$ orbitals ($\delta$) as a function of the
charge redistribution $\Delta q$ for Sr$_2$IrO$_4$ and Ba$_2$IrO$_4$.
}
\label{fig:coordination}
\end{figure}

\noindent{\bf Discussion}\\
The exact $d$-level order is of fundamental importance in TM oxides, dictating for instance the symmetry
of the quasiparticle states in photoemission \cite{Ir214_Kim_fermi_arcs_2014, Sr2RhO4_kim_2006,
  RhIr214_vmk_14}, the sign of the single-ion anisotropy \cite{Os227_bogdanov_12} and the nature of the
magnetic ordering \cite{Os227_bogdanov_12, Ca2RuO4_terakura_04}.
In 214 iridates specifically, it determines the various isotropic as well as anisotropic contributions to
the magnetic exchange couplings \cite{IrO_kitaev_jackeli_09, Ir214_BHKim_2012, Ir214_perkins_14,
  Ba214_vmk_14}, the evolution of those magnetic interactions with strain \cite{Ir214_lupascu_14} and/or
pressure \cite{Sr2IrO4_xmcd_2012} and most likely the nature of the intriguing transition to a
nonmagnetic phase in Sr$_2$IrO$_4$ under high pressure \cite{Sr2IrO4_xmcd_2012}.
Having established that in Sr$_2$IrO$_4$ the $d$ levels are inverted and that in the closely related
Ba$_2$IrO$_4$ they are not raises the question what actually drives the inversion.
To address this, we performed an additional set of calculations, in which we change the charges around
the reference IrO$_6$ octahedron.
As a simple numerical experiment that preserves charge neutrality of the A$_2$IrO$_4$ system, we assigned
the 4 NN iridium sites (in-plane, see Fig.\,\ref{fig:coordination}) the charge
$Q_{\mathrm{TM}}\!-\!2\Delta q$ and the 8 closest A-site cations (out of plane) the valence
$Q_{\mathrm{A}}\!+\!\Delta q$.
In a fully ionic picture, $Q_{\mathrm{TM}}$ and $Q_{\mathrm{A}}$ are $4+$ and $2+$, respectively.
However, since in our calculations the NN TM and A sites are not modeled as just formal point charges
(see Methods) the actual valence states depart from their formal values, with larger ``deviations'' for
$Q_{\mathrm{TM}}$.
The way we introduce $\Delta q$ in the computations is therefore by appropriately modifying the nuclear
charge at the respective site.
For variable $\Delta q$, this interpolates linearly between nearby surroundings corresponding to $5d$ 214
layered perovskites (with $\Delta q\!=\!0$ and TM$^{4+}$, A$^{2+}$  formal valence states) and their
cuprate 214 equivalents (with $\Delta q\!=\!1$, TM$^{2+}$/A$^{3+}$ formal ionic charges and ``normal''
order of the TM $t_{\mathrm{2g}}$ and $e_{\mathrm{g}}$ levels \cite{CuO2_dd_hozoi_11}).

As is illustrated in Fig.~\ref{fig:coordination}(a), increasing $\Delta q$ amounts to moving positive
charge from the IrO$_2$ plane to the adjacent A-O layers.
The calculations show that upon moving charge in such a manner, the Ir $t_{\mathrm{2g}}$ splitting $\delta$ 
  increases, see Fig.~\ref{fig:coordination}(b).
In other words, this redistribution of charge counteracts the level inversion in Sr$_2$IrO$_4$ and
further increases the already positive $\delta$ in Ba$_2$IrO$_4$.
In Sr$_2$IrO$_4$ the cubic-like $j_{\mathrm{eff}}\!=\!1/2$ limit occurs for $\Delta q\!=\!0.22$.
This effect can easily be understood: placing more positive charge out of the IrO$_2$ plane stabilizes
the out-of-plane $t_{\mathrm{2g}}$ orbitals, corresponding to the ($yz$,$zx$) orbital doublet, and thus enhances
$\delta$.
One can also do the opposite and drive $\Delta q$ negative.
In this case more positive charge piles up in the IrO$_2$ plane, which one expects to lower the energy of
the $xy$ orbital singlet, thus enhancing the level inversion in Sr$_2$IrO$_4$.
This is indeed what happens, see Fig.~\ref{fig:coordination}b.
What is more, driving $\Delta q$ negative even causes a level inversion in Ba$_2$IrO$_4$, when $\Delta
q\!\lesssim \!-0.25$.
It is interesting to note that the slope of the $\delta$ versus $\Delta q$ lines in Sr$_2$IrO$_4$ is much
steeper than in Ba$_2$IrO$_4$, which is caused by the significantly smaller Ir--Ir distances in
Sr$_2$IrO$_4$.

From these test calculations it is clear that low-symmetry crystal fields associated to neighbors beyond
the first ligand coordination shell -- in particular, the highly charged Ir$^{4+}$ NN's -- counteract the
local tetragonal crystal field that is caused by the elongation of the IrO$_6$ octahedra in both
Sr$_2$IrO$_4$ and Ba$_2$IrO$_4$.
In the case of Ba$_2$IrO$_4$ the local distortion is still strong enough to overcome these longer-range
effects but in Sr$_2$IrO$_4$, with a slightly smaller tetragonal distortion, the longer-range
electrostatics wins, causing the observed level inversion.

While the role of the high ionic charge of in-plane ions has been earlier invoked in the tetravalent Ru
oxide compound Ca$_2$RuO$_4$ \cite{Ca2RuO4_terakura_04} and in mixed-valence manganites
\cite{LaSrMnO4_wu_11}, we here explicitly prove it by combined ESR measurements and many-body {\it ab
  initio} calculations on structurally and chemically simpler systems in which additional complications
arising from further distortions \cite{Ca2RuO4_terakura_04,Ca2RuO4_braden_1998} or the presence of
multiple TM valence states \cite{LaSrMnO4_wu_11} are excluded.
A reversed order of the Ir $t_{\mathrm{2g}}$ levels in Sr$_2$IrO$_4$ has been also indirectly implied by fits of
x-ray absorption \cite{Sr2IrO4_xmcd_2012} and x-ray magnetic scattering \cite{Sr2IrO4_NRMXD_fujiyama13,
  Boseggia13} spectra.
As a more direct and more sensitive experimental technique to such details of the valence electronic
structure and with back up from truly {\it ab initio} many-body calculations, ESR now provides
irrefutable evidence for such physics.
The numerical ``experiment'' outlined in Fig.\,\ref{fig:coordination} further shows that at the heart of
this effect is not the intersite exchange, as assumed in Ref.\,\onlinecite{Sr2IrO4_xmcd_2012}, and not
the $t_{\mathrm{2g}}$-$e_{\mathrm{g}}$ orbital hybridization invoked in Ref.\,\onlinecite{Sr2IrO4_NRMXD_fujiyama13}, but
basic interlayer electrostatics.

We have, in sum, provided an integrated picture on the $d$-level structure and magnetic anisotropies in
Sr$_2$IrO$_4$, a prototype spin-orbit driven magnetic insulator.
Both the single-site $\bar{\bar{\vec g}}$ tensor and intersite effective exchange interactions are
analyzed in detail.
To access the latter, we build on an earlier computational scheme for deriving intersite matrix elements
in mixed-valence spin-orbit coupled systems \cite{g_bolvin_07}.
While the ratio $D$/$J$ of the antisymmetric Dzyaloshinskii-Moriya and isotropic Heisenberg couplings is
remarkably large in Sr$_2$IrO$_4$ and concurs with an in-plane rotation pattern of the Ir magnetic
moments that follows nearly rigidly the staggered rotation of the IrO$_6$ octahedra
\cite{Boseggia13,Ye13}, the most prominent symmetric anisotropic terms are according to the quantum
chemistry data in-plane, perpendicular to the Ir-Ir links.
The structure of the $\bar{\bar{\vec g}}$ tensor, as measured by ESR and computed with first-principles
electronic-structure methods, is such that $g_{\parallel}\!=\!g_c\!>\!g_{\perp}\!=\!g_a\!=\!g_b$ and
distinctly indicates a negative tetragonal-like splitting of the Ir $t_{\mathrm{2g}}$ levels, in spite of sizable
positive tetragonal distortions in Sr$_2$IrO$_4$.
We further observe that a much stronger tetragonal distortion in Ba$_2$IrO$_4$ renders the tetragonal
$d$-level splitting positive and $g_{\parallel}\!<\!g_{\perp}$.
The interesting situation arises that nevertheless the magnitude of the Ir $t_{\mathrm{2g}}$ splitting is largest
in Sr$_2$IrO$_4$.
The $d$-level inversion in Sr$_2$IrO$_4$ and the surprisingly small splitting in Ba$_2$IrO$_4$ have to do
with the way the positive ionic charge is distributed between adjacent Ir$^{4+}$O$_2$ and A$^{2+}$O
layers, having in contrast to the 214 cuprate superconductors, for example, more positive charge in the
TM-O planes.
This almost compensates the ``local'' tetragonal field arising from the $z$-axis elongation of the
IrO$_6$ octahedra in Ba$_2$IrO$_4$ and overcompensate it in Sr$_2$IrO$_4$.

The subtle interplay between local distortions of the O ligand cage and additional uniaxial fields
associated with the anisotropic extended surroundings opens new perspectives on strain
\cite{Ir214_lupascu_14} and pressure \cite{Sr2IrO4_xmcd_2012} experiments in square-lattice iridates, for
example, in connection to the spin-flop transition earlier predicted in Sr$_2$IrO$_4$
\cite{IrO_kitaev_jackeli_09,Ir214_perkins_14}.
It also opens up the perspective of manipulating this way the $d$-level ordering in oxide
heterostructures with highly charged, trivalent and tetravalent, species.
Compounds with tetravalent species within the TM-O$_2$ layers, in particular, given the nonpolar
character of the quasi 2D sheets, provide ideal playgrounds to explore the mechanism of $d$-level
ordering pointed out here since that will not be hindered by ``interface'' charge redistribution and
structural reconstruction occuring in polar heterostructures from polar discontinuities
\cite{polar_layers_ohtomo_04,polar_layers_koster_08}.
A recent experimental realization of such mixed, tetravalent/divalent TM-oxide interfaces is for example
the SrRuO$_3$/NiO interface \cite{interf_RuNi_ning13}, one system that requires in this respect closer
theoretical examination.
\\

{\footnotesize

\noindent{\bf Methods}\\ 
\noindent{\bf Single-site magnetic properties.}
The $g$ factors were obtained by computations on clusters which contain one central IrO$_6$ octahedron,
the four NN IrO$_6$ octahedra and the nearby ten Sr ions.
The solid-state surroundings were modeled as a large array of point charges fitted to reproduce the
crystal Madelung field in the cluster region.
To obtain a clear picture on crystal-field effects and spin-orbit interactions at the central Ir site, we
cut off the magnetic couplings with the adjacent Ir ions by replacing the tetravalent open-shell $d^5$
NN's with tetravalent closed-shell Pt$^{4+}$ $t_{\mathrm{2g}}^6$ species.
This is a usual procedure in quantum chemistry studies on TM systems, see, e.g.,
Refs.\,\onlinecite{qc_NNs_degraaf_99, Na2V2O5_hozoi_02, CuO2_dd_hozoi_11, NiO_domingo_12,
  Os227_bogdanov_12, SIA_Fe_maurice_2013}.
We used energy-consistent relativistic pseudopotentials and valence basis sets of quadruple-zeta quality
supplemented with $f$ polarization functions for the central Ir ion \cite{ECP_stoll_5d_09} and
all-electron triple-zeta basis sets for the six adjacent ligands \cite{BS_dunning_89}.
For the TM NN's, we applied energy-consistent relativistic pseudopotentials and triple-zeta basis
functions \cite{ECP_stoll_5d_09} along with minimal atomic-natural-orbital basis sets
\cite{BS_ano_pierloot_95} for the O's coordinating those TM sites but not shared with the central
octahedron.
The Sr and Ba species were modeled by divalent total-ion effective potentials supplemented with a single
$s$ function \cite{ECP_stoll_alkaline_85}.
All O $2p$ and metal $t_{\mathrm{2g}}$ electrons at the central octahedron were correlated in the MRCI
calculations.
The latter are performed with single and double substitutions with respect to the CASSCF reference (for
technicalities, see Refs.\,\onlinecite{CI_molpro_jcp_88,CI_molpro_tca_92}), which is
referred to as MRCISD. 
To separate the metal $5d$ and O $2p$ valence orbitals into different groups, i.e., central-octahedron
and adjacent-octahedra orbitals, we used the Pipek-Mezey localization module \cite{localization_PM}
available in  MOLPRO.
The computations with hypothetical $(Q_{\mathrm{TM}}\!-\!2\Delta q)$ and $(Q_{\mathrm{A}}\!+\!\Delta q)$
ionic charges at the TM and Sr/Ba sites next to the reference Ir ion were carried out as frozen-orbital
multideterminant calculations (also referred to as CASCI) with three Ir $t_{\mathrm{2g}}$ and five electrons in
the active space and orbitals optimized for $\Delta q\!=\!0$.

\begin{table}[!b]
\caption{ {\bf Ir $\bm{t_{\mathrm{2g}}}$ splittings and $\bm{g}$ factors for Sr$\bm{_2}$IrO$\bm{_4}$ and
    Ba$\bm{_2}$IrO$\bm{_4}$.}
  Results as obtained with the  ORCA program \cite{orca} are shown.
  Only the $^2T_{\mathrm{2g}}$($t_{\mathrm{2g}}^5$) states were included in the CASSCF optimization and in the spin-orbit
  treatment.
}
\label{gfactors_214_orca}
\begin{ruledtabular}
\begin{tabular}{lccc}
                      &$\delta_{t_{\mathrm{2g}}}$ (meV) &$g_{\perp}$ &$g_{\parallel}$ \\[0.03cm]
\hline
\\[-0.20cm]
Sr$_2$IrO$_4$\,:      &                        &            &    \\
CASSCF                &--127                   &1.66        &2.23\\
NEVPT2                &--199                   &1.55        &2.41\\
\\[-0.20cm]
Ba$_2$IrO$_4$\,:      &                        &            &    \\
CASSCF                &30                      &1.93        &1.74\\
NEVPT2                &70                      &2.01        &1.58\\
\end{tabular}
\end{ruledtabular}
\end{table}

\begin{table*}[!ht]
\caption{
{\bf Matrix elements of the {\em ab initio}
spin-orbit Hamiltonian.}
The latter is described by expression (\ref{QC_Hamm_2sites}).
Results of spin-orbit MRCI calculations are shown (meV).
The two-site singlet and (split) triplet states are labeled $\ket{\tilde s}$ and
\{$\ket{t_{x}}$,\,$\ket{t_{y}}$,\,$\ket{\tilde t_{z}}$\}, respectively.
Due to the antisymmetric exchange, $\ket{\tilde s}$ and $\ket{\tilde t_{z}}$ are admixtures of ``pure''
$\ket{0,0}$ and $\ket{1,0}$ spin functions.
}
\label{abinitioHam}
\begin{ruledtabular}
\begin{tabular}{rcccc}
$\mc{H}^{i,j}$
                 &$\ket{\tilde s}$ &$\ket{t_{x}}$ &$\ket{t_{y}}$ &$\ket{\tilde t_{z}}$ \\
$\bra{\tilde s}$ &$0$              &$0.2308i\,\mu_{\mathrm{B}} h_y$
                                                      &$-0.1768i\,\mu_{\mathrm{B}} h_x$
                                                                        &$0$ \\
$\bra{t_{x}}$
                 &$-0.2308i\,\mu_{\mathrm{B}} h_y$
                                   &$48.3328$         &$2.3083i\,\mu_{\mathrm{B}} h_z$
                                                                        &$-1.6854i\,\mu_{\mathrm{B}} h_y$ \\
$\bra{t_{y}}$
                 &$0.1768i\,\mu_{\mathrm{B}} h_x$
                                   &$-2.3083i\,\mu_{\mathrm{B}} h_z$
                                                      &$48.9626$        &$1.6266i\,\mu_{\mathrm{B}} h_x$ \\
$\bra{\tilde t_{z}}$
                 &$0$              &$1.6854i\,\mu_{\mathrm{B}} h_y$
                                                      &$-1.6266i\,\mu_{\mathrm{B}} h_x$
                                                                        &$49.0630$ \\
\end{tabular}
\end{ruledtabular}
\end{table*}

\begin{table*}[!t]
\caption{
{\bf Matrix elements of the effective spin Hamiltonian.}
The explicit form of the latter is given by (\ref{SpinHam_2sites}); for additional notations we use here,
see Table~\ref{smParameters}.
}
\label{smModHam}
\begin{ruledtabular}
\begin{tabular}{rcccc}
$\mc{H}^{i,j}_{\mathrm{eff}}$
                &$\ket{\tilde s}$ &$\ket{t_{x}}$ &$\ket{t_{y}}$ &$\ket{\tilde t_{z}}$ \\

$\bra{\tilde s}$
                & 0
                                  &$ i h_y\frac{ g_{yy}D-g_{xy} \left(\rfrac{\tilde{J}}{2}+J_{\!x\!+\!y}\right)}
      {\sqrt{\tilde{J} \left(\rfrac{\tilde{J}}{2}+J_{\!x\!+\!y}\right)}}$
                                                     &$ -i h_x\frac{ g_{xx}D+g_{xy}
    \left(\rfrac{\tilde{J}}{2}+J_{\!x\!+\!y}\right)}{ \sqrt{\tilde{J} \left(\rfrac{\tilde{J}}{2}+J_{\!x\!+\!y}\right)}}$
                                                                        & 0 \\

$\bra{t_{x}}$
                &$ -i h_y\frac{ ( g_{yy}D-g_{xy} \left(\rfrac{\tilde{J}}{2}+J_{\!x\!+\!y}\right)}{\sqrt{\tilde{J}
     \left(\rfrac{\tilde{J}}{2}+J_{\!x\!+\!y}\right)}}$
                                  &$ \frac{1}{2} \left(\rfrac{\tilde{J}}{2}+J_{\!x\!+\!y}-\Gamma_{\!x\!-\!y}+\Gamma_{z}
     \right)$
                                                     &$-i g_{zz} h_z$
                                                                        & $-i h_y\frac{g_{yy}D+g_{xy}
    \left(\rfrac{\tilde{J}}{2}-J_{\!x\!+\!y}\right)}{\sqrt{\tilde{J} \left(\rfrac{\tilde{J}}{2}-J_{\!x\!+\!y}\right)}}$ \\

$\bra{t_{y}}$
                &$i h_x \frac{g_{xx}D+g_{xy} \left(\rfrac{\tilde{J}}{2}+J_{\!x\!+\!y}\right)}{ \sqrt{\tilde{J}
      \left(\rfrac{\tilde{J}}{2}+J_{\!x\!+\!y}\right)}}$
                                  &$i g_{zz} h_z$ &$ \frac{1}{2} \left(\rfrac{\tilde{J}}{2}+J_{\!x\!+\!y}+
      \Gamma_{x\!-\!y}+\Gamma_{z}\right)$
                                                                      & $i h_x\frac{g_{xx}D-g_{xy}
      \left(\rfrac{\tilde{J}}{2}-J_{\!x\!+\!y}\right)}{\sqrt{\tilde{J} \left(\rfrac{\tilde{J}}{2}-J_{\!x\!+\!y}\right)}}$ \\

$\bra{\tilde t_{z}}$
                & 0
                                  &$i h_y \frac{g_{yy}D+g_{xy}
      \left(\rfrac{\tilde{J}}{2}-J_{\!x\!+\!y}\right)}{\sqrt{\tilde{J} \left(\rfrac{\tilde{J}}{2}-J_{\!x\!+\!y}\right)}}$

                                                     &$-i h_x \frac{g_{xx}D-g_{xy}
      \left(\rfrac{\tilde{J}}{2}-J_{\!x\!+\!y}\right)}{\sqrt{\tilde{J} \left(\rfrac{\tilde{J}}{2}-J_{\!x\!+\!y}\right)}}$
                                                                      &$ \rfrac{\tilde{J}}{2}$
\end{tabular}
\end{ruledtabular}
\end{table*}

The spin-orbit treatment was performed according to the procedure described in
Ref.~\onlinecite{SOC_molpro}.
In a first step, the scalar relativistic Hamiltonian is used to calculate correlated wave functions for a
finite number of low-lying states, either at the CASSCF or MRCI level.
In a second step, the spin-orbit part is added to the initial scalar relativistic Hamiltonian, matrix
elements of the aforementioned states are evaluated for this extended Hamiltonian and the resulting
matrix is finally diagonalized to yield spin-orbit coupled wave functions.

The $g$ factors were computed following the scheme proposed by Bolvin~\cite{g_bolvin_06} and Vancoillie
{\it et al.} \cite{g_vancoillie_07} (for alternative formulations, see, e.g.,
Ref.\,\onlinecite{g_chibotaru_12}).
For the KD GS $\left\{ \psi, \bar{\psi} \right\}$, the Abragam-Bleaney tensor
\cite{gerloch_mcmeeking_DaltonTrans_1975,book_abragam_bleaney} ${\bf G}={\bf g}{\bf g}^T$ can be written
in matrix form as
\begin{equation}
\begin{split}
G_{kl} & = 2\displaystyle\sum\limits_{u,v=\psi,\bar{\psi}} \bra{u} \hat{L}_k+g_{\mathrm{e}}\hat{S}_k \ket{v} \bra{v}
               \hat{L}_l+g_{\mathrm{e}}\hat{S}_l \ket{u}
         \\*
      &  = \displaystyle\sum\limits_{m=x,y,z} ( \Lambda_{km} + g_{\mathrm{e}} \Sigma_{km} ) (\Lambda_{lm} + g_{\mathrm{e}} \Sigma_{lm} ) \,,
\end{split}
\label{Gkl}
\end{equation}
where
\begin{equation}
\begin{aligned}
&\Lambda_{kx}\!=\!2Re\left[\bra{\bar{\psi}}\hat{L}_k\ket{\psi}\right] \,,
&\Sigma_{kx}\!=&\,2Re\left[\bra{\bar{\psi}}\hat{S}_k\ket{\psi}\right] \,,\\
&\Lambda_{ky}\!=\!2Im\left[\bra{\bar{\psi}}\hat{L}_k\ket{\psi}\right] \,,
&\Sigma_{ky}\!=&\,2Im\left[\bra{\bar{\psi}}\hat{S}_k\ket{\psi}\right] \,, \\
&\Lambda_{kz}\!=\!2\bra{\psi}\hat{L}_k\ket{\psi} \,,
&\Sigma_{kz}\!=&\,2\bra{\psi}\hat{S}_k\ket{\psi} \,.
\end{aligned}
\end{equation}
The ME's of $\hat{L}$ are here provided by  MOLPRO while those of $\hat{S}$ are derived using the
conventional expressions for the generalized Pauli matrices\,:
\begin{equation}
\label{PauliM}
\begin{split}
  (\hat{S}_z)_{MM^\prime} &= M \delta_{MM^\prime}\;,\\
  (\hat{S}_x)_{MM^\prime} &= \frac{1}{2}\sqrt{(S+M)(S-M+1)} \delta_{M-1,M^\prime} \\
                       &+ \frac{1}{2}\sqrt{(S-M)(S+M+1)} \delta_{M+1,M^\prime}\;,\\
  (\hat{S}_y)_{MM^\prime} &= -\frac{i}{2}\sqrt{(S+M)(S-M+1)} \delta_{M-1,M^\prime}  \\
                       &+ \frac{i}{2}\sqrt{(S-M)(S+M+1)} \delta_{M+1,M^\prime}.
\end{split}
\end{equation}
{\bf G} is next diagonalized and the $g$ factors are obtained as the possitive square roots of the three
eigenvalues.
The corresponding eigenvectors specify the rotation matrix to the main magnetic axes.
In our case, the magnetic $Z$ axis is along the crystallographic $c$ coordinate, while $X$ and $Y$ are
``degenerate'' and can be any two perpendicular directions in the $ab$ plane.

To cross-check the $g$-factor values computed with our subroutine, we further performed $g$-factor
calculations using the module available within the  ORCA quantum chemistry package \cite{orca}.
We applied all-electron DKH (Douglas-Kroll-Hess) basis sets of triple-zeta quality for the TM ions
\cite{Orca_Irbasis}, triple-zeta basis functions for the ligands of the central octahedron
\cite{BS_dunning_89} and double-zeta basis functions for additional O's at the NN octahedra
\cite{BS_dunning_89}.
Dynamical correlation effects were accounted for by $N$-electron valence-state second-order perturbation
theory (NEVPT2) \cite{NEVPT2_CPL_2001,NEVPT2_benchmark_2013}.
CASSCF and NEVPT2 results are listed in Table~\ref{gfactors_214_orca}, for both Sr$_2$IrO$_4$ and
Ba$_2$IrO$_4$.
It is seen that the data in  Tables \ref{gfactors_Sr$_2$IrO$_4$_qc} and \ref{gfactors_214_orca}
compare very well and indicate the same overall trends.

\noindent{\bf Superexchange interactions in Sr$\bm{_2}$IrO$\bm{_4}$.}
NN magnetic coupling constants were obtained for Sr$_2$IrO$_4$ by calculations on an embedded cluster
that includes two IrO$_6$ octahedra as magnetically active units.
As for the calculation of single-site magnetic properties, to accurately describe the charge distribution
in the immediate neighborhood, the adjacent IrO$_{6}$ octahedra (six) and the closest Sr$^{2+}$ ions (16)
were also incorporated in the actual cluster.
We used energy-consistent relativistic pseudopotentials along with quadruple-zeta basis sets for the
valence shells of the two magnetically active Ir ions~\cite{ECP_stoll_5d_09}, all-electron quintuple-zeta
basis sets for the bridging ligand \cite{BS_dunning_89} and triple-zeta basis functions for the other O's
associated with the two reference octahedra \cite{BS_dunning_89}.
We further employed polarization functions at the two central Ir sites and for the bridging anion, namely
2 Ir $f$ and 4 O $d$ functions \cite{ECP_stoll_5d_09,BS_dunning_89}.
Additional ions defining the NN octahedra, the nearby Sr$^{2+}$ species and the farther crystalline
surroundings were modeled as in the single-site study, see above.

\begin{table}[!ht]
\caption{
{\bf Notations used for anisotropic exchange coupling parameters in Sr$\bm{_2}$IrO$\bm{_4}$.}
The definitions on the left-hand side are applied in Table~\ref{smModHam}.
Other conventions presently employed in the literature are also shown for comparison.
}
\label{smParameters}
\begin{ruledtabular}
\begin{tabular}{c|c|c|c}
                &$J, \Gamma_{xx}\!=\!-\Gamma_{yy}\!-\!\Gamma_{zz}$           
                &$J_x,J_y,J_z$
                &$J,\delta\!J_{xy}, \delta\!J_z$ ($x$ bond)\\
                &(present study)
                &(Ref.\,\onlinecite{Ir214_perkins_14})
                &(Refs.\,\onlinecite{IrO_kitaev_jackeli_09,Ir214_perkins_14,Ir214_BHKim_2012})\\
\hline
$J_{\!x\!+\!y}$
                &$J-\rfrac{\Gamma_{\!zz}}{2}$
                                &$\rfrac{\left(J_x+J_y \right)}{2}$
                                                &$J+\rfrac{\delta\!J_{xy}}{2}$ \\
$\tilde{J}$
                &2$\sqrt{D^2+J^2_{\!x\!+\!y}}$
                                &2$\sqrt{D^2+J^2_{\!x\!+\!y}}$
                                                &2$\sqrt{D^2+J^2_{\!x\!+\!y}}$ \\
$\Gamma_{x\!-\!y}$
                &$\rfrac{\left(\Gamma_{xx}-\Gamma_{yy}\right)}{2}$
                                &$\rfrac{\left(J_x-J_y \right)}{2}$
                                                &$\left \{ \begin{matrix} \delta\!J_{xy} &{\mathrm{for}}~\bra{t_{x}}
                                                 \mc{H} \ket{t_{x}} \\  0 &{\mathrm{for}}~\bra{t_{y}} \mc{H}
                                                 \ket{t_{y}} \end{matrix} \right.$ \\
$\Gamma_z$
                &$\rfrac{3}{2}\Gamma_{zz}$
                                &$J_z-J_{\!x\!+\!y}$
                                                &$\delta\!J_z$
\end{tabular}
\end{ruledtabular}
\end{table}

For two adjacent magnetic sites, the ${^2T_{\mathrm{2g}}}(t_{\mathrm{2g}}^5)\!\otimes\!{^2T_{\mathrm{2g}}}(t_{\mathrm{2g}}^5)$ manifold
entails nine singlet and nine triplet states.
The CASSCF optimization was carried out for an average of these nine singlet and nine triplet
eigenfunctions of the scalar relativistic Hamiltonian $\mc{H}_{\rm SR}^{i,j}$.
In the subsequent MRCI treatment, only the Ir $t_{\mathrm{2g}}$ and the O $2p$ electrons at the bridging ligand
site were correlated.
Results in good agreement with the experimental data were recently obtained with this computational
approach for related 5$d^5$ iridates \cite{Ir113_bogdanov_12,Ir213_katukuri_14}.

Diagonalization of the Hamiltonian $\mc{H}_{\rm SR}^{i,j} + \mc{H}_{\rm SO}^{i,j}$ in the basis of the
lowest nine singlet and nine triplet states provides a total of 36 spin-orbit-coupled
$\ket{j_{\rm eff}^i,j_{\rm eff}^j}$
eigenfunctions, namely, four $\ket{\rfrac{1}{2},\rfrac{1}{2}}$, eight $\ket{\rfrac{1}{2},\rfrac{3}{2}}$,
eight $\ket{\rfrac{3}{2},\rfrac{1}{2}}$ and sixteen $\ket{\rfrac{3}{2},\rfrac{3}{2}}$ states.
In the simplest picture, the lowest four $\ket{\rfrac{1}{2},\rfrac{1}{2}}$ roots imply either singlet or
triplet coupling of the spin-orbit $j_{\rm eff}\!=\!1/2$ (or, more generally,
pseudospin $\tilde{S}\!=\!1/2$)
on-site objects and are separated from higher-lying states by a gap of $\gtrsim\!0.5$\,eV,
much larger than the strength of the intersite exchange.
It is this set of lowest four spin-orbit MRCI roots that we map onto the eigenstates of the effective
two-site (pseudo)spin Hamiltonian (\ref{SpinHam_2sites}).
The Zeeman interaction shows up on the quantum chemistry side as
$\mc{H}_{\rm Z}^{i,j}\!=\!\mu_{\rm B}(\vec{\hat{L}}^{i,j} + g_e\vec{\hat{S}}^{i,j})\cdot\vec{h}$,
where $\vec{\hat{L}}^{i,j}$ and $\vec{\hat{S}}^{i,j}$ are transformed to the spin-orbit-coupled basis using
the spin-orbit wave functions as unitary transformation matrix.
ME's of the {\it ab initio} model Hamiltonian
$\mc{H}_{\rm SR}^{i,j} + \mc{H}_{\rm SO}^{i,j} + \mc{H}_{\rm Z}^{i,j}$
are shown in Table~\ref{abinitioHam}.
Diagonal components show the energies of the zero-field states, while the off-diagonal ME's describe
the coupling to magnetic field.

For the experimentally determined crystal structure of Sr$_2$IrO$_4$ \cite{Sr214_crawford94}, the
two-octahedra [Ir$_2$O$_{11}$] cluster displays $C_{2v}$ symmetry.
Having the $x$ axis along the $\langle 110\rangle$ crystallographic direction \cite{Sr214_crawford94} and
$z\!\parallel\!c$, the effective anisotropic couplings read $\vec{D}=\left(0,0,D \right )$, \be
\label{smAnisoCopulings_x}
\begin{aligned}
\bar{\bar{\bs{\Gamma}}}^{(x)}\!=\!
\left(
        \begin{array}{ccc}
        \Gamma_{xx} & 0 & 0 \\
        0 & \Gamma_{yy} & 0 \\
        0 & 0 & \Gamma_{zz}
        \end{array}
\right)
\end{aligned}
\ee
for Ir-Ir links along $x$ and
\be
\label{smAnisoCopulings_y}
\begin{aligned}
\bar{\bar{\bs{\Gamma}}}^{(y)}\!=\!
\left(
        \begin{array}{ccc}
        \Gamma_{yy} & 0 & 0 \\
        0 & \Gamma_{xx}& 0 \\
        0 & 0 & \Gamma_{zz}
        \end{array}
\right)
\end{aligned}
\ee
for Ir-Ir links along $y$, 
with $\Gamma_{xx} +\Gamma_{yy} +\Gamma_{zz} =0$.
The uniform and staggered components of the $\bar{\bar{\vec g}}$ tensor take for individual Ir sites the
following form\,:
\be
\label{sm_2site_g}
\begin{aligned}
\bar{\bar{\vec g}}_i\!=&\,\bar{\bar{\vec g}}_i^{\mathrm{u}} \pm \bar{\bar{\vec g}}_i^{\mathrm{s}}\!=\!
\!\left(
        \begin{array}{ccc}
        g_{xx} & 0 & 0 \\
        0 & g_{yy} & 0 \\
        0 & 0 & g_{zz}
        \end{array}
\right)\! \pm
\!\left(
        \begin{array}{ccc}
        0 & g_{xy} & 0 \\
        g_{xy} & 0 & 0 \\
        0 & 0 & 0
        \end{array}
\right).
\end{aligned}
\ee
$g_{zz}\!=\!g_{\parallel}$ while $g_{xx}$ and $g_{yy}$ are directly related to $g_{\perp}$ but not
restricted to be equal due to the lower symmetry of the two-octahedra cluster as compared to the IrO$_6$
unit.

To solve now the actual problem, we need to transform the effective spin Hamiltonian
(\ref{SpinHam_2sites}) to the same form as the {\it ab initio} Hamiltonian shown in
Table~\ref{abinitioHam}, i.e., diagonal in zero magnetic field.
The result of such a transformation is shown in Table~\ref{smModHam}.
Direct correspondence between homologous ME's in the two arrays yields a set of eight independent
equations that finally allow to derive hard values for all effective coupling constants that enter
expression (\ref{SpinHam_2sites}).
While the results for the intersite exchange interactions in (\ref{SpinHam_2sites}) and
(\ref{smAnisoCopulings_x},\ref{smAnisoCopulings_y}) are shown in Table \ref{2sites_results},
the $\bar{\bar{\vec g}}$-tensor data obtained from the two-octahedra calculations are
$g_{xx}\!=\!1.64$, $g_{yy}\!=\!1.70$, $g_{zz}\!=\!2.31$, and $g_{xy}\!=\!0.02$.
The way the additional parameters introduced in Table~\ref{smModHam} are defined is explained in
Table~\ref{smParameters}.
The intermediate steps followed to arrive to the matrix form provided in Table~\ref{smModHam} are
outlined in Supplementary Tables I--III and the Supplementary Methods.

\noindent{\bf Spin-wave calculations.}
For understanding all details of the ESR spectrum, we carried out a spin-wave analysis using 
the Hamiltonian (\ref{SpinHam_2sites}) and a site-factorized variational GS wave
function
\be
|\Psi\rangle=\prod_{\rm u.c.}|\psi_A\rangle|\psi_B\rangle\;,
\ee
where for each sublattice index $L\!\in\!\left\{A,B\right\}$,
\be
|\psi_L\rangle=\cos\frac{\alpha_L}{2}|\uparrow\rangle+e^{i\varphi_L} \sin\frac{\alpha_L}{2}|\downarrow\rangle\;.
\ee
For the 2D unit cell displayed in Fig.\,\ref{fig:symm}(a), $D_{2d}$ symmetry is considered.

The spin components of the GS configuration depend on the $\alpha_L$ and $\varphi_L$ variational
parameters as
\be
S^x_L = \frac{1}{2}\sin\alpha_L\cos\varphi_L\;,
\ee
\be
S^y_L = \frac{1}{2}\sin\alpha_L\sin\varphi_L\;,
\ee
\be
S^z_L = \frac{1}{2}\cos\alpha_L\;.
\ee

For magnetic fields parallel to the $c$ axis, the GS energy only depends on the parameters
$\alpha_A=\alpha_B=\alpha$ and $\varphi_A-\varphi_B=\delta\varphi$.
$\delta\varphi$ is the angle between neighboring in-plane spins and $\alpha$ describes how much the spins
are tilted away from the $c$ axis.
For $\alpha=\pi/2$ the spins are lying within the $ab$ plane, while $\alpha=0$ corresponds to the fully
polarized high-magnetic-field case.
The GS energy
\begin{eqnarray}
E_0&=&(J+\Gamma_{\!zz})\cos\alpha^2-g_\|h_\|\cos\alpha\nonumber\\
   &+&((J-\frac{1}{2}\Gamma_{\!zz})\cos\delta\varphi-D\sin\delta\varphi)\sin\alpha^2
\end{eqnarray}
is minimized when
\be
\delta\varphi=-\arccos\frac{2J-\Gamma_{\!zz}}{\tilde J}\;,
\ee
\be
\alpha=\arccos\frac{g_\| h_\|}{2J+2\Gamma_{\!zz}+\tilde J}\;.
\ee

For zero field we find that $\alpha=0$, the spins are confined to the $ab$ plane (i.e., to the IrO$_2$
layer) and the angle $\delta\varphi$ is controlled by the strength of $D$.
$\delta\varphi$ is not affected by fields along the $c$ axis while $\alpha$ changes smoothly from $\pi/2$
to $0$ with increasing the field strength.

The two magnons characteristic for spin-1/2 antiferromagnets are related to states orthogonal to
$|\psi_A\rangle$ and $|\psi_B\rangle$.
The $4\times4$ Hamiltonian defining these magnons can be derived by using the well known
Holstein-Primakoff approach and diagonalized through a Bogoliubov transformation.
For $h\|c$, the two spin-wave modes are the gapless $\nu_{\|1}=0$ Goldstone mode, corresponding to U(1)
symmetry breaking, and the gapped mode
given by (\ref{in_pl_ESR}).
When the magnetic field lies in the $ab$ plane, the $z$ component of Ir spins remains zero (with
$\alpha=\pi/2$) and the GS energy only depends on the $\delta\varphi$ angle\,:
\begin{eqnarray}
E_0&=&(J+\frac{\Gamma_{\!zz}}{2})\cos\delta\varphi-g_xh_x\cos\frac{\delta\varphi}{2}-D\sin\delta\varphi\nonumber \,.\\
\label{eq:e0_hx}
\end{eqnarray}
For simplicity, we select $x$ for the direction of the magnetic field.
The result is, however, independent of how this choice is made as there is no anisotropy within the $ab$
plane.

An infinitesimally small in-plane field fixes the direction of uniform magnetization as
$m=-\cos\frac{\delta\varphi}{2}$.
At finite field both modes thus become gapped.
To determine the uniform magnetization $m$ one needs to minimize Eq.~(\ref{eq:e0_hx}), 
which to leading order in magnetic field leads to Eq.\;(\ref{eq:m}).
The first term of Eq.\,(\ref{eq:m}) corresponds to the zero-field moment which arises due to the canting
induced by the DM interaction. This ferromagnetic order-parameter is further enhanced in finite $h_\bot$
field. As long as $h_\bot$ is small Eq.\,(\ref{eq:m}) remains valid.

As discussed in the main text, in low fields, $m$ can be approximated by its field independent value
\cite{Bahr14}.
Using the quantum chemically derived interaction parameters (see Table\,\ref{2sites_results}), we then
find $m\!\approx\!0.12\mu_{\mathrm{B}}$, in good agreement with recent experiments
\cite{Sr2IrO4_bjkim_09,Sr214_cao98}.
Yet the zero-field gap comes out too large as compared to experiment.
A good fit can nevertheless be reached with $J$, $D$ and $g$ values as obtained in the MRCI treatment and
by increasing $\Gamma_{\!zz}$ from 0.42 to 0.98 meV.

\noindent{\bf Acknowledgments}\\
We thank G.\,Jackeli, G.\,Khaliullin, P.\,Fulde, and H.\,Takagi for fruitful discussions.
N.\,A.\,B. and L.\,H. acknowledge financial support from the Erasmus Mundus Programme of the European
Union and the German Research Foundation (Deutsche Forschungsgemeinschaft, DFG), respectively.
Experimental work was in part supported through the DFG project KA 1694/8-1.
\\

\noindent{ Correspondence should be addressed to N.A.\,B. (email: n.bogdanov@ifw-dresden.de) or
    L.\,H. (email: l.hozoi@ifw-dresden.de).}
\\

\noindent{\bf Authors Contribution}\\
V.\,K. conducted the ESR experiments.
~N.\,A.\,B. and V.\,M.\,K. carried out the {\it ab initio} quantum chemistry calculations and subsequent
mapping of the {\it ab initio} data onto the effective spin Hamiltonian, with assistance from L.\,H.,
V.\,Y., J.\,R. and V.\,K.
~J.\,R. performed the spin-wave analysis.
~L.\,H., V.\,K., J.\,v.d.B. and B.\,B. designed the project.
~L.\,H., J.\,v.d.B., V.\,K., N.\,A.\,B., V.\,M.\,K. and J.\,R. wrote the paper, with contributions from
all other coauthors.
}
 

\begin{thebibliography}{70}%
\makeatletter
\providecommand \@ifxundefined [1]{%
 \@ifx{#1\undefined}
}%
\providecommand \@ifnum [1]{%
 \ifnum #1\expandafter \@firstoftwo
 \else \expandafter \@secondoftwo
 \fi
}%
\providecommand \@ifx [1]{%
 \ifx #1\expandafter \@firstoftwo
 \else \expandafter \@secondoftwo
 \fi
}%
\providecommand \natexlab [1]{#1}%
\providecommand \enquote  [1]{``#1''}%
\providecommand \bibnamefont  [1]{#1}%
\providecommand \bibfnamefont [1]{#1}%
\providecommand \citenamefont [1]{#1}%
\providecommand \href@noop [0]{\@secondoftwo}%
\providecommand \href [0]{\begingroup \@sanitize@url \@href}%
\providecommand \@href[1]{\@@startlink{#1}\@@href}%
\providecommand \@@href[1]{\endgroup#1\@@endlink}%
\providecommand \@sanitize@url [0]{\catcode `\\12\catcode `\$12\catcode
  `\&12\catcode `\#12\catcode `\^12\catcode `\_12\catcode `\%12\relax}%
\providecommand \@@startlink[1]{}%
\providecommand \@@endlink[0]{}%
\providecommand \url  [0]{\begingroup\@sanitize@url \@url }%
\providecommand \@url [1]{\endgroup\@href {#1}{\urlprefix }}%
\providecommand \urlprefix  [0]{URL }%
\providecommand \Eprint [0]{\href }%
\providecommand \doibase [0]{http://dx.doi.org/}%
\providecommand \selectlanguage [0]{\@gobble}%
\providecommand \bibinfo  [0]{\@secondoftwo}%
\providecommand \bibfield  [0]{\@secondoftwo}%
\providecommand \translation [1]{[#1]}%
\providecommand \BibitemOpen [0]{}%
\providecommand \bibitemStop [0]{}%
\providecommand \bibitemNoStop [0]{.\EOS\space}%
\providecommand \EOS [0]{\spacefactor3000\relax}%
\providecommand \BibitemShut  [1]{\csname bibitem#1\endcsname}%
\let\auto@bib@innerbib\@empty
\bibitem [{\citenamefont {Heber}(2009)}]{heterostruct_heber_2009}%
  \BibitemOpen
  \bibfield  {author} {\bibinfo {author} {\bibfnamefont {J.}~\bibnamefont
  {Heber}},\ }\bibfield  {title} {\enquote {\bibinfo {title} {Materials
  science: Enter the oxides},}\ }\href@noop {} {\bibfield  {journal} {\bibinfo
  {journal} {Nature}\ }\textbf {\bibinfo {volume} {459}},\ \bibinfo {pages}
  {28} (\bibinfo {year} {2009})}\BibitemShut {NoStop}%
\bibitem [{\citenamefont {Mannhart}\ and\ \citenamefont
  {Schlom}(2010)}]{heterostruct_mannhart_2010}%
  \BibitemOpen
  \bibfield  {author} {\bibinfo {author} {\bibfnamefont {J.}~\bibnamefont
  {Mannhart}}\ and\ \bibinfo {author} {\bibfnamefont {D.~G.}\ \bibnamefont
  {Schlom}},\ }\bibfield  {title} {\enquote {\bibinfo {title} {Oxide interfaces
  -- an opportunity for electronics},}\ }\href@noop {} {\bibfield  {journal}
  {\bibinfo  {journal} {Science}\ }\textbf {\bibinfo {volume} {327}},\ \bibinfo
  {pages} {1607} (\bibinfo {year} {2010})}\BibitemShut {NoStop}%
\bibitem [{\citenamefont {Chakhalian}\ \emph {et~al.}(2007)\citenamefont
  {Chakhalian}, \citenamefont {Freeland}, \citenamefont {Habermeier},
  \citenamefont {Cristiani}, \citenamefont {Khaliullin}, \citenamefont {van
  Veenendaal},\ and\ \citenamefont {Keimer}}]{OrbRec_chakhalian_2007}%
  \BibitemOpen
  \bibfield  {author} {\bibinfo {author} {\bibfnamefont {J.}~\bibnamefont
  {Chakhalian}}, \bibinfo {author} {\bibfnamefont {J.~W.}\ \bibnamefont
  {Freeland}}, \bibinfo {author} {\bibfnamefont {H.-U.}\ \bibnamefont
  {Habermeier}}, \bibinfo {author} {\bibfnamefont {G.}~\bibnamefont
  {Cristiani}}, \bibinfo {author} {\bibfnamefont {G.}~\bibnamefont
  {Khaliullin}}, \bibinfo {author} {\bibfnamefont {M.}~\bibnamefont {van
  Veenendaal}}, \ and\ \bibinfo {author} {\bibfnamefont {B.}~\bibnamefont
  {Keimer}},\ }\bibfield  {title} {\enquote {\bibinfo {title} {Orbital
  reconstruction and covalent bonding at an oxide interface},}\ }\href@noop {}
  {\bibfield  {journal} {\bibinfo  {journal} {Science}\ }\textbf {\bibinfo
  {volume} {318}},\ \bibinfo {pages} {1114} (\bibinfo {year}
  {2007})}\BibitemShut {NoStop}%
\bibitem [{\citenamefont {Benckiser}\ \emph {et~al.}(2011)\citenamefont
  {Benckiser}, \citenamefont {Haverkort}, \citenamefont {Br\"{u}ck},
  \citenamefont {Goering}, \citenamefont {Macke}, \citenamefont
  {Fra{\~n}\'{o}}, \citenamefont {Yang}, \citenamefont {Andersen},
  \citenamefont {Cristiani}, \citenamefont {Habermeier}, \citenamefont {Boris},
  \citenamefont {Zegkinoglou}, \citenamefont {Wochner}, \citenamefont {Kim},
  \citenamefont {Hinkov},\ and\ \citenamefont
  {Keimer}}]{OrbRec_benckiser_2011}%
  \BibitemOpen
  \bibfield  {author} {\bibinfo {author} {\bibfnamefont {E.}~\bibnamefont
  {Benckiser}}, \bibinfo {author} {\bibfnamefont {M.~W.}\ \bibnamefont
  {Haverkort}}, \bibinfo {author} {\bibfnamefont {S.}~\bibnamefont
  {Br\"{u}ck}}, \bibinfo {author} {\bibfnamefont {E.}~\bibnamefont {Goering}},
  \bibinfo {author} {\bibfnamefont {S.}~\bibnamefont {Macke}}, \bibinfo
  {author} {\bibfnamefont {A.}~\bibnamefont {Fra{\~n}\'{o}}}, \bibinfo {author}
  {\bibfnamefont {X.}~\bibnamefont {Yang}}, \bibinfo {author} {\bibfnamefont
  {O.~K.}\ \bibnamefont {Andersen}}, \bibinfo {author} {\bibfnamefont
  {G.}~\bibnamefont {Cristiani}}, \bibinfo {author} {\bibfnamefont {H.-U.}\
  \bibnamefont {Habermeier}}, \bibinfo {author} {\bibfnamefont {A.~V.}\
  \bibnamefont {Boris}}, \bibinfo {author} {\bibfnamefont {I.}~\bibnamefont
  {Zegkinoglou}}, \bibinfo {author} {\bibfnamefont {P.}~\bibnamefont
  {Wochner}}, \bibinfo {author} {\bibfnamefont {H.-J.}\ \bibnamefont {Kim}},
  \bibinfo {author} {\bibfnamefont {V.}~\bibnamefont {Hinkov}}, \ and\ \bibinfo
  {author} {\bibfnamefont {B.}~\bibnamefont {Keimer}},\ }\bibfield  {title}
  {\enquote {\bibinfo {title} {Orbital reflectometry of oxide
  heterostructures},}\ }\href@noop {} {\bibfield  {journal} {\bibinfo
  {journal} {Nature Materials}\ }\textbf {\bibinfo {volume} {10}},\ \bibinfo
  {pages} {189} (\bibinfo {year} {2011})}\BibitemShut {NoStop}%
\bibitem [{\citenamefont {Hwang}\ \emph {et~al.}(2012)\citenamefont {Hwang},
  \citenamefont {Iwasa}, \citenamefont {Kawasaki}, \citenamefont {Keimer},
  \citenamefont {Nagaosa},\ and\ \citenamefont {Tokura}}]{OrbRec_tokura_2012}%
  \BibitemOpen
  \bibfield  {author} {\bibinfo {author} {\bibfnamefont {H.~Y.}\ \bibnamefont
  {Hwang}}, \bibinfo {author} {\bibfnamefont {Y.}~\bibnamefont {Iwasa}},
  \bibinfo {author} {\bibfnamefont {M.}~\bibnamefont {Kawasaki}}, \bibinfo
  {author} {\bibfnamefont {B.}~\bibnamefont {Keimer}}, \bibinfo {author}
  {\bibfnamefont {N.}~\bibnamefont {Nagaosa}}, \ and\ \bibinfo {author}
  {\bibfnamefont {Y.}~\bibnamefont {Tokura}},\ }\bibfield  {title} {\enquote
  {\bibinfo {title} {Emergent phenomena at oxide interfaces},}\ }\href@noop {}
  {\bibfield  {journal} {\bibinfo  {journal} {Nature Materials}\ }\textbf
  {\bibinfo {volume} {11}},\ \bibinfo {pages} {103} (\bibinfo {year}
  {2012})}\BibitemShut {NoStop}%
\bibitem [{\citenamefont {{Leggett}}(2006)}]{Leggett06}%
  \BibitemOpen
  \bibfield  {author} {\bibinfo {author} {\bibfnamefont {A.~J.}\ \bibnamefont
  {{Leggett}}},\ }\bibfield  {title} {\enquote {\bibinfo {title} {{What DO we
  know about high T$_{c}$?}}}\ }\href {\doibase 10.1038/nphys254} {\bibfield
  {journal} {\bibinfo  {journal} {Nature Physics}\ }\textbf {\bibinfo {volume}
  {2}},\ \bibinfo {pages} {134--136} (\bibinfo {year} {2006})}\BibitemShut
  {NoStop}%
\bibitem [{\citenamefont {Kim}\ \emph {et~al.}(2008)\citenamefont {Kim},
  \citenamefont {Jin}, \citenamefont {Moon}, \citenamefont {Kim}, \citenamefont
  {Park}, \citenamefont {Leem}, \citenamefont {Yu}, \citenamefont {Noh},
  \citenamefont {Kim}, \citenamefont {Oh}, \citenamefont {Park}, \citenamefont
  {Durairaj}, \citenamefont {Cao},\ and\ \citenamefont
  {Rotenberg}}]{Sr2IrO4_bjkim_08}%
  \BibitemOpen
  \bibfield  {author} {\bibinfo {author} {\bibfnamefont {B.~J.}\ \bibnamefont
  {Kim}}, \bibinfo {author} {\bibfnamefont {H.}~\bibnamefont {Jin}}, \bibinfo
  {author} {\bibfnamefont {S.~J.}\ \bibnamefont {Moon}}, \bibinfo {author}
  {\bibfnamefont {J.-Y.}\ \bibnamefont {Kim}}, \bibinfo {author} {\bibfnamefont
  {B.-G.}\ \bibnamefont {Park}}, \bibinfo {author} {\bibfnamefont {C.~S.}\
  \bibnamefont {Leem}}, \bibinfo {author} {\bibfnamefont {J.}~\bibnamefont
  {Yu}}, \bibinfo {author} {\bibfnamefont {T.~W.}\ \bibnamefont {Noh}},
  \bibinfo {author} {\bibfnamefont {C.}~\bibnamefont {Kim}}, \bibinfo {author}
  {\bibfnamefont {S.-J.}\ \bibnamefont {Oh}}, \bibinfo {author} {\bibfnamefont
  {J.-H.}\ \bibnamefont {Park}}, \bibinfo {author} {\bibfnamefont
  {V.}~\bibnamefont {Durairaj}}, \bibinfo {author} {\bibfnamefont
  {G.}~\bibnamefont {Cao}}, \ and\ \bibinfo {author} {\bibfnamefont
  {E.}~\bibnamefont {Rotenberg}},\ }\bibfield  {title} {\enquote {\bibinfo
  {title} {Novel ${J}_{\mathrm{eff}}=1/2$ {M}ott state induced by relativistic
  spin-orbit coupling in $\mathrm{{S}r_{2}{I}r{O}_{4}}$},}\ }\href@noop {}
  {\bibfield  {journal} {\bibinfo  {journal} {Phys. Rev. Lett.}\ }\textbf
  {\bibinfo {volume} {101}},\ \bibinfo {pages} {076402} (\bibinfo {year}
  {2008})}\BibitemShut {NoStop}%
\bibitem [{\citenamefont {Kim}\ \emph {et~al.}(2009)\citenamefont {Kim},
  \citenamefont {Ohsumi}, \citenamefont {Komesu}, \citenamefont {Sakai},
  \citenamefont {Morita}, \citenamefont {Takagi},\ and\ \citenamefont
  {Arima}}]{Sr2IrO4_bjkim_09}%
  \BibitemOpen
  \bibfield  {author} {\bibinfo {author} {\bibfnamefont {B.~J.}\ \bibnamefont
  {Kim}}, \bibinfo {author} {\bibfnamefont {H.}~\bibnamefont {Ohsumi}},
  \bibinfo {author} {\bibfnamefont {T.}~\bibnamefont {Komesu}}, \bibinfo
  {author} {\bibfnamefont {S.}~\bibnamefont {Sakai}}, \bibinfo {author}
  {\bibfnamefont {T.}~\bibnamefont {Morita}}, \bibinfo {author} {\bibfnamefont
  {H.}~\bibnamefont {Takagi}}, \ and\ \bibinfo {author} {\bibfnamefont
  {T.}~\bibnamefont {Arima}},\ }\bibfield  {title} {\enquote {\bibinfo {title}
  {Phase-sensitive observation of a spin-orbital {M}ott state in
  $\mathrm{{S}r_{2}{I}r{O}_{4}}$},}\ }\href@noop {} {\bibfield  {journal}
  {\bibinfo  {journal} {Science}\ }\textbf {\bibinfo {volume} {323}},\ \bibinfo
  {pages} {1329} (\bibinfo {year} {2009})}\BibitemShut {NoStop}%
\bibitem [{\citenamefont {Kim}\ \emph {et~al.}(2012{\natexlab{a}})\citenamefont
  {Kim}, \citenamefont {Casa}, \citenamefont {Upton}, \citenamefont {Gog},
  \citenamefont {Kim}, \citenamefont {Mitchell}, \citenamefont {van
  Veenendaal}, \citenamefont {Daghofer}, \citenamefont {van~den Brink},
  \citenamefont {Khaliullin},\ and\ \citenamefont {Kim}}]{Ir214_jkim_2012}%
  \BibitemOpen
  \bibfield  {author} {\bibinfo {author} {\bibfnamefont {J.}~\bibnamefont
  {Kim}}, \bibinfo {author} {\bibfnamefont {D.}~\bibnamefont {Casa}}, \bibinfo
  {author} {\bibfnamefont {M.~H.}\ \bibnamefont {Upton}}, \bibinfo {author}
  {\bibfnamefont {T.}~\bibnamefont {Gog}}, \bibinfo {author} {\bibfnamefont
  {Y.-J.}\ \bibnamefont {Kim}}, \bibinfo {author} {\bibfnamefont {J.~F.}\
  \bibnamefont {Mitchell}}, \bibinfo {author} {\bibfnamefont {M.}~\bibnamefont
  {van Veenendaal}}, \bibinfo {author} {\bibfnamefont {M.}~\bibnamefont
  {Daghofer}}, \bibinfo {author} {\bibfnamefont {J.}~\bibnamefont {van~den
  Brink}}, \bibinfo {author} {\bibfnamefont {G.}~\bibnamefont {Khaliullin}}, \
  and\ \bibinfo {author} {\bibfnamefont {B.~J.}\ \bibnamefont {Kim}},\
  }\bibfield  {title} {\enquote {\bibinfo {title} {Magnetic excitation spectra
  of $\mathrm{{S}r_{2}{I}r{O}_{4}}$ probed by resonant inelastic x-ray
  scattering: Establishing links to cuprate superconductors},}\ }\href@noop {}
  {\bibfield  {journal} {\bibinfo  {journal} {Phys. Rev. Lett.}\ }\textbf
  {\bibinfo {volume} {108}},\ \bibinfo {pages} {177003} (\bibinfo {year}
  {2012}{\natexlab{a}})}\BibitemShut {NoStop}%
\bibitem [{\citenamefont {Kim}\ \emph {et~al.}(2014{\natexlab{a}})\citenamefont
  {Kim}, \citenamefont {Krupin}, \citenamefont {Denlinger}, \citenamefont
  {Bostwick}, \citenamefont {Rotenberg}, \citenamefont {Zhao}, \citenamefont
  {Mitchell}, \citenamefont {Allen},\ and\ \citenamefont
  {Kim}}]{Ir214_Kim_fermi_arcs_2014}%
  \BibitemOpen
  \bibfield  {author} {\bibinfo {author} {\bibfnamefont {Y.~K.}\ \bibnamefont
  {Kim}}, \bibinfo {author} {\bibfnamefont {O.}~\bibnamefont {Krupin}},
  \bibinfo {author} {\bibfnamefont {J.~D.}\ \bibnamefont {Denlinger}}, \bibinfo
  {author} {\bibfnamefont {A.}~\bibnamefont {Bostwick}}, \bibinfo {author}
  {\bibfnamefont {E.}~\bibnamefont {Rotenberg}}, \bibinfo {author}
  {\bibfnamefont {Q.}~\bibnamefont {Zhao}}, \bibinfo {author} {\bibfnamefont
  {J.~F.}\ \bibnamefont {Mitchell}}, \bibinfo {author} {\bibfnamefont {J.~W.}\
  \bibnamefont {Allen}}, \ and\ \bibinfo {author} {\bibfnamefont {B.~J.}\
  \bibnamefont {Kim}},\ }\bibfield  {title} {\enquote {\bibinfo {title}
  {{F}ermi arcs in a doped pseudospin-1/2 {H}eisenberg antiferromagnet},}\
  }\href {\doibase 10.1126/science.1251151} {\bibfield  {journal} {\bibinfo
  {journal} {Science}\ }\textbf {\bibinfo {volume} {345}},\ \bibinfo {pages}
  {187} (\bibinfo {year} {2014}{\natexlab{a}})}\BibitemShut {NoStop}%
\bibitem [{\citenamefont {Kim}\ \emph {et~al.}(2014{\natexlab{b}})\citenamefont
  {Kim}, \citenamefont {Daghofer}, \citenamefont {Said}, \citenamefont {Gog},
  \citenamefont {van~den Brink}, \citenamefont {Khaliullin},\ and\
  \citenamefont {Kim}}]{Ir214_Kim_excitonQP_2014}%
  \BibitemOpen
  \bibfield  {author} {\bibinfo {author} {\bibfnamefont {J.}~\bibnamefont
  {Kim}}, \bibinfo {author} {\bibfnamefont {M.}~\bibnamefont {Daghofer}},
  \bibinfo {author} {\bibfnamefont {A.~H.}\ \bibnamefont {Said}}, \bibinfo
  {author} {\bibfnamefont {T.}~\bibnamefont {Gog}}, \bibinfo {author}
  {\bibfnamefont {J.}~\bibnamefont {van~den Brink}}, \bibinfo {author}
  {\bibfnamefont {G.}~\bibnamefont {Khaliullin}}, \ and\ \bibinfo {author}
  {\bibfnamefont {B.~J.}\ \bibnamefont {Kim}},\ }\bibfield  {title} {\enquote
  {\bibinfo {title} {Excitonic quasiparticles in a spin-orbit {M}ott
  insulator},}\ }\href@noop {} {\bibfield  {journal} {\bibinfo  {journal} {Nat.
  Commun.}\ }\textbf {\bibinfo {volume} {5}},\ \bibinfo {pages} {4453}
  (\bibinfo {year} {2014}{\natexlab{b}})}\BibitemShut {NoStop}%
\bibitem [{\citenamefont {Jackeli}\ and\ \citenamefont
  {Khaliullin}(2009)}]{IrO_kitaev_jackeli_09}%
  \BibitemOpen
  \bibfield  {author} {\bibinfo {author} {\bibfnamefont {G.}~\bibnamefont
  {Jackeli}}\ and\ \bibinfo {author} {\bibfnamefont {G.}~\bibnamefont
  {Khaliullin}},\ }\bibfield  {title} {\enquote {\bibinfo {title} {{M}ott
  insulators in the strong spin-orbit coupling limit: From {H}eisenberg to a
  quantum compass and {K}itaev models},}\ }\href@noop {} {\bibfield  {journal}
  {\bibinfo  {journal} {Phys. Rev. Lett.}\ }\textbf {\bibinfo {volume} {102}},\
  \bibinfo {pages} {017205} (\bibinfo {year} {2009})}\BibitemShut {NoStop}%
\bibitem [{\citenamefont {Abragam}\ and\ \citenamefont
  {Bleaney}(1970)}]{book_abragam_bleaney}%
  \BibitemOpen
  \bibfield  {author} {\bibinfo {author} {\bibfnamefont {A.}~\bibnamefont
  {Abragam}}\ and\ \bibinfo {author} {\bibfnamefont {B.}~\bibnamefont
  {Bleaney}},\ }\href@noop {} {\emph {\bibinfo {title} {Electron Paramagnetic
  Resonance of Transition Ions}}}\ (\bibinfo  {publisher} {Clarendon Press,
  Oxford},\ \bibinfo {year} {1970})\BibitemShut {NoStop}%
\bibitem [{\citenamefont {Boseggia}\ \emph
  {et~al.}(2013{\natexlab{a}})\citenamefont {Boseggia}, \citenamefont
  {Springell}, \citenamefont {Walker}, \citenamefont {R{\o}nnow}, \citenamefont
  {R\"{u}egg}, \citenamefont {Okabe}, \citenamefont {Isobe}, \citenamefont
  {Perry}, \citenamefont {Collins},\ and\ \citenamefont
  {McMorrow}}]{Ba2IrO4_xmcd_2013}%
  \BibitemOpen
  \bibfield  {author} {\bibinfo {author} {\bibfnamefont {S.}~\bibnamefont
  {Boseggia}}, \bibinfo {author} {\bibfnamefont {R.}~\bibnamefont {Springell}},
  \bibinfo {author} {\bibfnamefont {H.~C.}\ \bibnamefont {Walker}}, \bibinfo
  {author} {\bibfnamefont {H.~M.}\ \bibnamefont {R{\o}nnow}}, \bibinfo {author}
  {\bibfnamefont {Ch.}\ \bibnamefont {R\"{u}egg}}, \bibinfo {author}
  {\bibfnamefont {H.}~\bibnamefont {Okabe}}, \bibinfo {author} {\bibfnamefont
  {M.}~\bibnamefont {Isobe}}, \bibinfo {author} {\bibfnamefont {R.~S.}\
  \bibnamefont {Perry}}, \bibinfo {author} {\bibfnamefont {S.~P.}\ \bibnamefont
  {Collins}}, \ and\ \bibinfo {author} {\bibfnamefont {D.~F.}\ \bibnamefont
  {McMorrow}},\ }\bibfield  {title} {\enquote {\bibinfo {title} {Robustness of
  basal-plane antiferromagnetic order and the ${J}_{\mathrm{eff}}\mathbf{=}1/2$
  state in single-layer iridate spin-orbit {M}ott insulators},}\ }\href@noop {}
  {\bibfield  {journal} {\bibinfo  {journal} {Phys. Rev. Lett.}\ }\textbf
  {\bibinfo {volume} {110}},\ \bibinfo {pages} {117207} (\bibinfo {year}
  {2013}{\natexlab{a}})}\BibitemShut {NoStop}%
\bibitem [{\citenamefont {Thornley}(1968)}]{SOC_d5_thornley68}%
  \BibitemOpen
  \bibfield  {author} {\bibinfo {author} {\bibfnamefont {J.~H.~M.}\
  \bibnamefont {Thornley}},\ }\bibfield  {title} {\enquote {\bibinfo {title}
  {The magnetic properties of $(\mathrm{{I}r{X}}_6)^{2-}$ complexes},}\ }\href
  {http://stacks.iop.org/0022-3719/1/i=4/a=321} {\bibfield  {journal} {\bibinfo
   {journal} {J. Phys. C (Proc. Phys. Soc.)}\ }\textbf {\bibinfo {volume}
  {1}},\ \bibinfo {pages} {1024} (\bibinfo {year} {1968})}\BibitemShut
  {NoStop}%
\bibitem [{\citenamefont {Hill}(1972)}]{SOC_d5_hill71}%
  \BibitemOpen
  \bibfield  {author} {\bibinfo {author} {\bibfnamefont {N.~J.}\ \bibnamefont
  {Hill}},\ }\bibfield  {title} {\enquote {\bibinfo {title} {Electron
  paramagnetic resonance of os\-mi\-um-do\-ped trichlorotris
  (diethylphenylphosphine) rho\-dium ($\mathrm{III}$)},}\ }\href {\doibase
  10.1039/F29726800427} {\bibfield  {journal} {\bibinfo  {journal} {J. Chem.
  Soc.{,} Faraday Trans. 2}\ }\textbf {\bibinfo {volume} {68}},\ \bibinfo
  {pages} {427--434} (\bibinfo {year} {1972})}\BibitemShut {NoStop}%
\bibitem [{\citenamefont {Andlauer}\ \emph {et~al.}(1976)\citenamefont
  {Andlauer}, \citenamefont {Schneider},\ and\ \citenamefont
  {Tolksdorf}}]{SOC_d5_andlauer76}%
  \BibitemOpen
  \bibfield  {author} {\bibinfo {author} {\bibfnamefont {B.}~\bibnamefont
  {Andlauer}}, \bibinfo {author} {\bibfnamefont {J.}~\bibnamefont {Schneider}},
  \ and\ \bibinfo {author} {\bibfnamefont {W.}~\bibnamefont {Tolksdorf}},\
  }\bibfield  {title} {\enquote {\bibinfo {title} {Optical absorption,
  fluorescence, and electron spin resonance of $\mathrm{Ir}^{4+}$ on octahedral
  sites in $\mathrm{Y_3Ga_5O_{12}}$},}\ }\href@noop {} {\bibfield  {journal}
  {\bibinfo  {journal} {Phys. Stat. Sol. B}\ }\textbf {\bibinfo {volume}
  {73}},\ \bibinfo {pages} {533} (\bibinfo {year} {1976})}\BibitemShut
  {NoStop}%
\bibitem [{\citenamefont {Stebler}\ \emph {et~al.}(1984)\citenamefont
  {Stebler}, \citenamefont {Ammeter}, \citenamefont {F\"{u}rholz},\ and\
  \citenamefont {Ludi}}]{SOC_d5_stebler84}%
  \BibitemOpen
  \bibfield  {author} {\bibinfo {author} {\bibfnamefont {A.}~\bibnamefont
  {Stebler}}, \bibinfo {author} {\bibfnamefont {J.~M.}\ \bibnamefont
  {Ammeter}}, \bibinfo {author} {\bibfnamefont {U.}~\bibnamefont
  {F\"{u}rholz}}, \ and\ \bibinfo {author} {\bibfnamefont {A.}~\bibnamefont
  {Ludi}},\ }\bibfield  {title} {\enquote {\bibinfo {title} {Electronic
  properties of hexaaquaruthenium ($\mathrm{III}$): $\mathrm{EPR}$ and optical
  spectra of $\mathrm{Ru(H_2O)_6^{3+}}$ in alum crystals},}\ }\href {\doibase
  10.1021/ic00182a033} {\bibfield  {journal} {\bibinfo  {journal} {Inorg.
  Chem.}\ }\textbf {\bibinfo {volume} {23}},\ \bibinfo {pages} {2764} (\bibinfo
  {year} {1984})}\BibitemShut {NoStop}%
\bibitem [{\citenamefont {Haskel}\ \emph {et~al.}(2012)\citenamefont {Haskel},
  \citenamefont {Fabbris}, \citenamefont {Zhernenkov}, \citenamefont {Kong},
  \citenamefont {Jin}, \citenamefont {Cao},\ and\ \citenamefont {van
  Veenendaal}}]{Sr2IrO4_xmcd_2012}%
  \BibitemOpen
  \bibfield  {author} {\bibinfo {author} {\bibfnamefont {D.}~\bibnamefont
  {Haskel}}, \bibinfo {author} {\bibfnamefont {G.}~\bibnamefont {Fabbris}},
  \bibinfo {author} {\bibfnamefont {M.}~\bibnamefont {Zhernenkov}}, \bibinfo
  {author} {\bibfnamefont {P.~P.}\ \bibnamefont {Kong}}, \bibinfo {author}
  {\bibfnamefont {C.}~\bibnamefont {Jin}}, \bibinfo {author} {\bibfnamefont
  {G.}~\bibnamefont {Cao}}, \ and\ \bibinfo {author} {\bibfnamefont
  {M.}~\bibnamefont {van Veenendaal}},\ }\bibfield  {title} {\enquote {\bibinfo
  {title} {Pressure tuning of the spin-orbit coupled ground state in
  {S}r$_2${I}r{O}$_4$},}\ }\href@noop {} {\bibfield  {journal} {\bibinfo
  {journal} {Phys. Rev. Lett.}\ }\textbf {\bibinfo {volume} {109}},\ \bibinfo
  {pages} {027204} (\bibinfo {year} {2012})}\BibitemShut {NoStop}%
\bibitem [{\citenamefont {Liu}\ \emph {et~al.}(2012)\citenamefont {Liu},
  \citenamefont {Katukuri}, \citenamefont {Hozoi}, \citenamefont {Yin},
  \citenamefont {Dean}, \citenamefont {Upton}, \citenamefont {Kim},
  \citenamefont {Casa}, \citenamefont {Said}, \citenamefont {Gog},
  \citenamefont {Qi}, \citenamefont {Cao}, \citenamefont {Tsvelik},
  \citenamefont {van~den Brink},\ and\ \citenamefont
  {Hill}}]{Ir3116_rixs_liu_2012}%
  \BibitemOpen
  \bibfield  {author} {\bibinfo {author} {\bibfnamefont {X.}~\bibnamefont
  {Liu}}, \bibinfo {author} {\bibfnamefont {Vamshi~M.}\ \bibnamefont
  {Katukuri}}, \bibinfo {author} {\bibfnamefont {L.}~\bibnamefont {Hozoi}},
  \bibinfo {author} {\bibfnamefont {Wei-Guo}\ \bibnamefont {Yin}}, \bibinfo
  {author} {\bibfnamefont {M.~P.~M.}\ \bibnamefont {Dean}}, \bibinfo {author}
  {\bibfnamefont {M.~H.}\ \bibnamefont {Upton}}, \bibinfo {author}
  {\bibfnamefont {Jungho}\ \bibnamefont {Kim}}, \bibinfo {author}
  {\bibfnamefont {D.}~\bibnamefont {Casa}}, \bibinfo {author} {\bibfnamefont
  {A.}~\bibnamefont {Said}}, \bibinfo {author} {\bibfnamefont {T.}~\bibnamefont
  {Gog}}, \bibinfo {author} {\bibfnamefont {T.~F.}\ \bibnamefont {Qi}},
  \bibinfo {author} {\bibfnamefont {G.}~\bibnamefont {Cao}}, \bibinfo {author}
  {\bibfnamefont {A.~M.}\ \bibnamefont {Tsvelik}}, \bibinfo {author}
  {\bibfnamefont {Jeroen}\ \bibnamefont {van~den Brink}}, \ and\ \bibinfo
  {author} {\bibfnamefont {J.~P.}\ \bibnamefont {Hill}},\ }\bibfield  {title}
  {\enquote {\bibinfo {title} {Testing the validity of the strong
  spin-orbit-coupling limit for octahedrally coordinated iridate compounds in a
  model system {S}r$_{3}${C}u{I}r{O}$_{6}$},}\ }\href@noop {} {\bibfield
  {journal} {\bibinfo  {journal} {Phys. Rev. Lett.}\ }\textbf {\bibinfo
  {volume} {109}},\ \bibinfo {pages} {157401} (\bibinfo {year}
  {2012})}\BibitemShut {NoStop}%
\bibitem [{\citenamefont {Bogdanov}\ \emph {et~al.}(2012)\citenamefont
  {Bogdanov}, \citenamefont {Katukuri}, \citenamefont {Stoll}, \citenamefont
  {van~den Brink},\ and\ \citenamefont {Hozoi}}]{Ir113_bogdanov_12}%
  \BibitemOpen
  \bibfield  {author} {\bibinfo {author} {\bibfnamefont {N.~A.}\ \bibnamefont
  {Bogdanov}}, \bibinfo {author} {\bibfnamefont {V.~M.}\ \bibnamefont
  {Katukuri}}, \bibinfo {author} {\bibfnamefont {H.}~\bibnamefont {Stoll}},
  \bibinfo {author} {\bibfnamefont {J.}~\bibnamefont {van~den Brink}}, \ and\
  \bibinfo {author} {\bibfnamefont {L.}~\bibnamefont {Hozoi}},\ }\bibfield
  {title} {\enquote {\bibinfo {title} {Post-perovskite {C}a{I}r{O}${}_{3}$: A
  $j=1/2$ quasi-one-dimensional antiferromagnet},}\ }\href@noop {} {\bibfield
  {journal} {\bibinfo  {journal} {Phys. Rev. B}\ }\textbf {\bibinfo {volume}
  {85}},\ \bibinfo {pages} {235147} (\bibinfo {year} {2012})}\BibitemShut
  {NoStop}%
\bibitem [{\citenamefont {Hozoi}\ \emph {et~al.}(2014)\citenamefont {Hozoi},
  \citenamefont {Gretarsson}, \citenamefont {Clancy}, \citenamefont {Jeon},
  \citenamefont {Lee}, \citenamefont {Kim}, \citenamefont {Yushankhai},
  \citenamefont {Fulde}, \citenamefont {Casa}, \citenamefont {Gog},
  \citenamefont {Kim}, \citenamefont {Said}, \citenamefont {Upton},
  \citenamefont {Kim},\ and\ \citenamefont {van~den Brink}}]{Ir227_hozoi_14}%
  \BibitemOpen
  \bibfield  {author} {\bibinfo {author} {\bibfnamefont {L.}~\bibnamefont
  {Hozoi}}, \bibinfo {author} {\bibfnamefont {H.}~\bibnamefont {Gretarsson}},
  \bibinfo {author} {\bibfnamefont {J.~P.}\ \bibnamefont {Clancy}}, \bibinfo
  {author} {\bibfnamefont {B.-G.}\ \bibnamefont {Jeon}}, \bibinfo {author}
  {\bibfnamefont {B.}~\bibnamefont {Lee}}, \bibinfo {author} {\bibfnamefont
  {K.~H.}\ \bibnamefont {Kim}}, \bibinfo {author} {\bibfnamefont
  {V.}~\bibnamefont {Yushankhai}}, \bibinfo {author} {\bibfnamefont
  {P.}~\bibnamefont {Fulde}}, \bibinfo {author} {\bibfnamefont
  {D.}~\bibnamefont {Casa}}, \bibinfo {author} {\bibfnamefont {T.}~\bibnamefont
  {Gog}}, \bibinfo {author} {\bibfnamefont {J.}~\bibnamefont {Kim}}, \bibinfo
  {author} {\bibfnamefont {A.~H.}\ \bibnamefont {Said}}, \bibinfo {author}
  {\bibfnamefont {M.~H.}\ \bibnamefont {Upton}}, \bibinfo {author}
  {\bibfnamefont {Y.~J.}\ \bibnamefont {Kim}}, \ and\ \bibinfo {author}
  {\bibfnamefont {J.}~\bibnamefont {van~den Brink}},\ }\bibfield  {title}
  {\enquote {\bibinfo {title} {Longer-range lattice anisotropy strongly
  competing with spin-orbit interactions in pyrochlore iridates},}\ }\href@noop
  {} {\bibfield  {journal} {\bibinfo  {journal} {Phys. Rev. B}\ }\textbf
  {\bibinfo {volume} {89}},\ \bibinfo {pages} {115111} (\bibinfo {year}
  {2014})}\BibitemShut {NoStop}%
\bibitem [{\citenamefont {{Crawford}}\ \emph {et~al.}(1994)\citenamefont
  {{Crawford}}, \citenamefont {{Subramanian}}, \citenamefont {{Harlow}},
  \citenamefont {{Fernandez-Baca}}, \citenamefont {{Wang}},\ and\ \citenamefont
  {{Johnston}}}]{Sr214_crawford94}%
  \BibitemOpen
  \bibfield  {author} {\bibinfo {author} {\bibfnamefont {M.~K.}\ \bibnamefont
  {{Crawford}}}, \bibinfo {author} {\bibfnamefont {M.~A.}\ \bibnamefont
  {{Subramanian}}}, \bibinfo {author} {\bibfnamefont {R.~L.}\ \bibnamefont
  {{Harlow}}}, \bibinfo {author} {\bibfnamefont {J.~A.}\ \bibnamefont
  {{Fernandez-Baca}}}, \bibinfo {author} {\bibfnamefont {Z.~R.}\ \bibnamefont
  {{Wang}}}, \ and\ \bibinfo {author} {\bibfnamefont {D.~C.}\ \bibnamefont
  {{Johnston}}},\ }\bibfield  {title} {\enquote {\bibinfo {title} {Structural
  and magnetic studies of {S}r$_{2}${I}r{O}$_{4}$},}\ }\href@noop {} {\bibfield
   {journal} {\bibinfo  {journal} {Phys. Rev. B}\ }\textbf {\bibinfo {volume}
  {49}},\ \bibinfo {pages} {9198} (\bibinfo {year} {1994})}\BibitemShut
  {NoStop}%
\bibitem [{\citenamefont {Ballhausen}(1962)}]{book_ballhausen_62}%
  \BibitemOpen
  \bibfield  {author} {\bibinfo {author} {\bibfnamefont {C.~J.}\ \bibnamefont
  {Ballhausen}},\ }\href@noop {} {\emph {\bibinfo {title} {Introduction to
  Ligand Field Theory}}}\ (\bibinfo  {publisher} {McGraw-Hill, New York},\
  \bibinfo {year} {1962})\BibitemShut {NoStop}%
\bibitem [{\citenamefont {Oshikawa}\ and\ \citenamefont
  {Affleck}(1997)}]{stagg_g_oshikawa_97}%
  \BibitemOpen
  \bibfield  {author} {\bibinfo {author} {\bibfnamefont {M.}~\bibnamefont
  {Oshikawa}}\ and\ \bibinfo {author} {\bibfnamefont {I.}~\bibnamefont
  {Affleck}},\ }\bibfield  {title} {\enquote {\bibinfo {title} {Field-induced
  gap in $\mathit{S}=1/2$ antiferromagnetic chains},}\ }\href@noop {}
  {\bibfield  {journal} {\bibinfo  {journal} {Phys. Rev. Lett.}\ }\textbf
  {\bibinfo {volume} {79}},\ \bibinfo {pages} {2883} (\bibinfo {year}
  {1997})}\BibitemShut {NoStop}%
\bibitem [{\citenamefont {Bahr}\ \emph {et~al.}(2014)\citenamefont {Bahr},
  \citenamefont {Alfonsov}, \citenamefont {Jackeli}, \citenamefont
  {Khaliullin}, \citenamefont {Matsumoto}, \citenamefont {Takayama},
  \citenamefont {Takagi}, \citenamefont {B\"uchner},\ and\ \citenamefont
  {Kataev}}]{Bahr14}%
  \BibitemOpen
  \bibfield  {author} {\bibinfo {author} {\bibfnamefont {S.}~\bibnamefont
  {Bahr}}, \bibinfo {author} {\bibfnamefont {A.}~\bibnamefont {Alfonsov}},
  \bibinfo {author} {\bibfnamefont {G.}~\bibnamefont {Jackeli}}, \bibinfo
  {author} {\bibfnamefont {G.}~\bibnamefont {Khaliullin}}, \bibinfo {author}
  {\bibfnamefont {A.}~\bibnamefont {Matsumoto}}, \bibinfo {author}
  {\bibfnamefont {T.}~\bibnamefont {Takayama}}, \bibinfo {author}
  {\bibfnamefont {H.}~\bibnamefont {Takagi}}, \bibinfo {author} {\bibfnamefont
  {B.}~\bibnamefont {B\"uchner}}, \ and\ \bibinfo {author} {\bibfnamefont
  {V.}~\bibnamefont {Kataev}},\ }\bibfield  {title} {\enquote {\bibinfo {title}
  {Low-energy magnetic excitations in the spin-orbital {M}ott insulator
  {Sr}$_{2}${I}r{O}$_{4}$},}\ }\href {\doibase 10.1103/PhysRevB.89.180401}
  {\bibfield  {journal} {\bibinfo  {journal} {Phys. Rev. B}\ }\textbf {\bibinfo
  {volume} {89}},\ \bibinfo {pages} {180401} (\bibinfo {year}
  {2014})}\BibitemShut {NoStop}%
\bibitem [{\citenamefont {Bolvin}(2006)}]{g_bolvin_06}%
  \BibitemOpen
  \bibfield  {author} {\bibinfo {author} {\bibfnamefont {H.}~\bibnamefont
  {Bolvin}},\ }\bibfield  {title} {\enquote {\bibinfo {title} {An alternative
  approach to the $g$-matrix: Theory and applications},}\ }\href@noop {}
  {\bibfield  {journal} {\bibinfo  {journal} {ChemPhysChem}\ }\textbf {\bibinfo
  {volume} {7}},\ \bibinfo {pages} {1575} (\bibinfo {year} {2006})}\BibitemShut
  {NoStop}%
\bibitem [{\citenamefont {Vancoillie}\ \emph {et~al.}(2007)\citenamefont
  {Vancoillie}, \citenamefont {Malmqvist},\ and\ \citenamefont
  {Pierloot}}]{g_vancoillie_07}%
  \BibitemOpen
  \bibfield  {author} {\bibinfo {author} {\bibfnamefont {S.}~\bibnamefont
  {Vancoillie}}, \bibinfo {author} {\bibfnamefont {P.~{\AA}.}\ \bibnamefont
  {Malmqvist}}, \ and\ \bibinfo {author} {\bibfnamefont {K.}~\bibnamefont
  {Pierloot}},\ }\bibfield  {title} {\enquote {\bibinfo {title} {Calculation of
  {EPR} $g$ tensors for transition-metal complexes based on
  multiconfigurational perturbation theory ({CASPT2})},}\ }\href@noop {}
  {\bibfield  {journal} {\bibinfo  {journal} {ChemPhysChem}\ }\textbf {\bibinfo
  {volume} {8}},\ \bibinfo {pages} {1803} (\bibinfo {year} {2007})}\BibitemShut
  {NoStop}%
\bibitem [{\citenamefont {Helgaker}\ \emph {et~al.}(2000)\citenamefont
  {Helgaker}, \citenamefont {J{\o}rgensen},\ and\ \citenamefont
  {Olsen}}]{book_QC_00}%
  \BibitemOpen
  \bibfield  {author} {\bibinfo {author} {\bibfnamefont {T.}~\bibnamefont
  {Helgaker}}, \bibinfo {author} {\bibfnamefont {P.}~\bibnamefont
  {J{\o}rgensen}}, \ and\ \bibinfo {author} {\bibfnamefont {J.}~\bibnamefont
  {Olsen}},\ }\href@noop {} {\emph {\bibinfo {title} {{Molecular
  Electronic-Structure Theory}}}}\ (\bibinfo  {publisher} {Wiley, Chichester},\
  \bibinfo {year} {2000})\BibitemShut {NoStop}%
\bibitem [{\citenamefont {Berning}\ \emph {et~al.}(2000)\citenamefont
  {Berning}, \citenamefont {Schweizer}, \citenamefont {Werner}, \citenamefont
  {Knowles},\ and\ \citenamefont {Palmieri}}]{SOC_molpro}%
  \BibitemOpen
  \bibfield  {author} {\bibinfo {author} {\bibfnamefont {A.}~\bibnamefont
  {Berning}}, \bibinfo {author} {\bibfnamefont {M.}~\bibnamefont {Schweizer}},
  \bibinfo {author} {\bibfnamefont {H.-J.}\ \bibnamefont {Werner}}, \bibinfo
  {author} {\bibfnamefont {P.~J.}\ \bibnamefont {Knowles}}, \ and\ \bibinfo
  {author} {\bibfnamefont {P.}~\bibnamefont {Palmieri}},\ }\bibfield  {title}
  {\enquote {\bibinfo {title} {Spin-orbit matrix elements for internally
  contracted multireference configuration interaction wavefunctions},}\
  }\href@noop {} {\bibfield  {journal} {\bibinfo  {journal} {Mol. Phys.}\
  }\textbf {\bibinfo {volume} {98}},\ \bibinfo {pages} {1823--1833} (\bibinfo
  {year} {2000})}\BibitemShut {NoStop}%
\bibitem [{\citenamefont {Werner}\ \emph {et~al.}()\citenamefont {Werner},
  \citenamefont {Knowles}, \citenamefont {Knizia}, \citenamefont {Manby},\ and\
  \citenamefont {Sch\"{u}tz}}]{molpro_brief}%
  \BibitemOpen
  \bibfield  {author} {\bibinfo {author} {\bibfnamefont {H.-J.}\ \bibnamefont
  {Werner}}, \bibinfo {author} {\bibfnamefont {P.~J.}\ \bibnamefont {Knowles}},
  \bibinfo {author} {\bibfnamefont {G.}~\bibnamefont {Knizia}}, \bibinfo
  {author} {\bibfnamefont {F.~R.}\ \bibnamefont {Manby}}, \ and\ \bibinfo
  {author} {\bibfnamefont {M.}~\bibnamefont {Sch\"{u}tz}},\ }\href@noop {}
  {}\bibinfo {note} {{MOLPRO} 2012, University of Cardiff, see
  http://www.molpro.net}\BibitemShut {NoStop}%
\bibitem [{\citenamefont {Katukuri}\ \emph
  {et~al.}(2014{\natexlab{a}})\citenamefont {Katukuri}, \citenamefont
  {Roszeitis}, \citenamefont {Yushankhai}, \citenamefont {Mitrushchenkov},
  \citenamefont {Stoll}, \citenamefont {van Veenendaal}, \citenamefont {Fulde},
  \citenamefont {van~den Brink},\ and\ \citenamefont {Hozoi}}]{RhIr214_vmk_14}%
  \BibitemOpen
  \bibfield  {author} {\bibinfo {author} {\bibfnamefont {V.~M.}\ \bibnamefont
  {Katukuri}}, \bibinfo {author} {\bibfnamefont {K.}~\bibnamefont {Roszeitis}},
  \bibinfo {author} {\bibfnamefont {V.}~\bibnamefont {Yushankhai}}, \bibinfo
  {author} {\bibfnamefont {A.}~\bibnamefont {Mitrushchenkov}}, \bibinfo
  {author} {\bibfnamefont {H.}~\bibnamefont {Stoll}}, \bibinfo {author}
  {\bibfnamefont {M.}~\bibnamefont {van Veenendaal}}, \bibinfo {author}
  {\bibfnamefont {P.}~\bibnamefont {Fulde}}, \bibinfo {author} {\bibfnamefont
  {J.}~\bibnamefont {van~den Brink}}, \ and\ \bibinfo {author} {\bibfnamefont
  {L.}~\bibnamefont {Hozoi}},\ }\bibfield  {title} {\enquote {\bibinfo {title}
  {Electronic structure of low-dimensional 4$d^5$ oxides: interplay of ligand
  distortions, overall lattice anisotropy, and spin-orbit interactions},}\
  }\href@noop {} {\bibfield  {journal} {\bibinfo  {journal} {Inorg. Chem.}\
  }\textbf {\bibinfo {volume} {53}},\ \bibinfo {pages} {4833} (\bibinfo {year}
  {2014}{\natexlab{a}})}\BibitemShut {NoStop}%
\bibitem [{\citenamefont {Okabe}\ \emph {et~al.}(2011)\citenamefont {Okabe},
  \citenamefont {Isobe}, \citenamefont {Takayama-Muromachi}, \citenamefont
  {Koda}, \citenamefont {Takeshita}, \citenamefont {Hiraishi}, \citenamefont
  {Miyazaki}, \citenamefont {Kadono}, \citenamefont {Miyake},\ and\
  \citenamefont {Akimitsu}}]{Ba214_okabe11}%
  \BibitemOpen
  \bibfield  {author} {\bibinfo {author} {\bibfnamefont {H.}~\bibnamefont
  {Okabe}}, \bibinfo {author} {\bibfnamefont {M.}~\bibnamefont {Isobe}},
  \bibinfo {author} {\bibfnamefont {E.}~\bibnamefont {Takayama-Muromachi}},
  \bibinfo {author} {\bibfnamefont {A.}~\bibnamefont {Koda}}, \bibinfo {author}
  {\bibfnamefont {S.}~\bibnamefont {Takeshita}}, \bibinfo {author}
  {\bibfnamefont {M.}~\bibnamefont {Hiraishi}}, \bibinfo {author}
  {\bibfnamefont {M.}~\bibnamefont {Miyazaki}}, \bibinfo {author}
  {\bibfnamefont {R.}~\bibnamefont {Kadono}}, \bibinfo {author} {\bibfnamefont
  {Y.}~\bibnamefont {Miyake}}, \ and\ \bibinfo {author} {\bibfnamefont
  {J.}~\bibnamefont {Akimitsu}},\ }\bibfield  {title} {\enquote {\bibinfo
  {title} {{B}a$_2${I}r{O}$_4$: A spin-orbit {M}ott insulating
  quasi-two-dimensional antiferromagnet},}\ }\href@noop {} {\bibfield
  {journal} {\bibinfo  {journal} {Phys. Rev. B}\ }\textbf {\bibinfo {volume}
  {83}},\ \bibinfo {pages} {155118} (\bibinfo {year} {2011})}\BibitemShut
  {NoStop}%
\bibitem [{\citenamefont {{Subramanian}}\ \emph {et~al.}(1994)\citenamefont
  {{Subramanian}}, \citenamefont {{Crawford}}, \citenamefont {{Harlow}},
  \citenamefont {Ami}, \citenamefont {{Fernandez-Baca}}, \citenamefont
  {{Wang}},\ and\ \citenamefont {{Johnston}}}]{Sr214_johnston94}%
  \BibitemOpen
  \bibfield  {author} {\bibinfo {author} {\bibfnamefont {M.~A.}\ \bibnamefont
  {{Subramanian}}}, \bibinfo {author} {\bibfnamefont {M.~K.}\ \bibnamefont
  {{Crawford}}}, \bibinfo {author} {\bibfnamefont {R.~L.}\ \bibnamefont
  {{Harlow}}}, \bibinfo {author} {\bibfnamefont {T.}~\bibnamefont {Ami}},
  \bibinfo {author} {\bibfnamefont {J.~A.}\ \bibnamefont {{Fernandez-Baca}}},
  \bibinfo {author} {\bibfnamefont {Z.~R.}\ \bibnamefont {{Wang}}}, \ and\
  \bibinfo {author} {\bibfnamefont {D.~C.}\ \bibnamefont {{Johnston}}},\
  }\bibfield  {title} {\enquote {\bibinfo {title} {{S}r$_2${R}h{O}$_4$ and
  {S}r2{I}r{O}$_4$: Structural and magnetic studies of 4$d$ and 5$d$ transition
  metal analogs of {La$_2$CuO$_4$}},}\ }\href@noop {} {\bibfield  {journal}
  {\bibinfo  {journal} {Physica C}\ }\textbf {\bibinfo {volume} {235-240}},\
  \bibinfo {pages} {743} (\bibinfo {year} {1994})}\BibitemShut {NoStop}%
\bibitem [{\citenamefont {Kim}\ \emph {et~al.}(2012{\natexlab{b}})\citenamefont
  {Kim}, \citenamefont {Khaliullin},\ and\ \citenamefont
  {Min}}]{Ir214_BHKim_2012}%
  \BibitemOpen
  \bibfield  {author} {\bibinfo {author} {\bibfnamefont {B.~H.}\ \bibnamefont
  {Kim}}, \bibinfo {author} {\bibfnamefont {G.}~\bibnamefont {Khaliullin}}, \
  and\ \bibinfo {author} {\bibfnamefont {B.~I.}\ \bibnamefont {Min}},\
  }\bibfield  {title} {\enquote {\bibinfo {title} {Magnetic couplings, optical
  spectra, and spin-orbit exciton in 5$d$ electron {M}ott insulator
  {Sr$_2$IrO$_4$}},}\ }\href@noop {} {\bibfield  {journal} {\bibinfo  {journal}
  {Phys. Rev. Lett.}\ }\textbf {\bibinfo {volume} {109}},\ \bibinfo {pages}
  {167205} (\bibinfo {year} {2012}{\natexlab{b}})}\BibitemShut {NoStop}%
\bibitem [{\citenamefont {Perkins}\ \emph {et~al.}(2014)\citenamefont
  {Perkins}, \citenamefont {Sizyuk},\ and\ \citenamefont
  {W{\"o}lfle}}]{Ir214_perkins_14}%
  \BibitemOpen
  \bibfield  {author} {\bibinfo {author} {\bibfnamefont {N.~B.}\ \bibnamefont
  {Perkins}}, \bibinfo {author} {\bibfnamefont {Y.}~\bibnamefont {Sizyuk}}, \
  and\ \bibinfo {author} {\bibfnamefont {P.}~\bibnamefont {W{\"o}lfle}},\
  }\bibfield  {title} {\enquote {\bibinfo {title} {{Interplay of many-body and
  single-particle interactions in iridates and rhodates}},}\ }\href@noop {}
  {\bibfield  {journal} {\bibinfo  {journal} {Phys. Rev. B}\ }\textbf {\bibinfo
  {volume} {89}},\ \bibinfo {pages} {035143} (\bibinfo {year}
  {2014})}\BibitemShut {NoStop}%
\bibitem [{\citenamefont {Boseggia}\ \emph
  {et~al.}(2013{\natexlab{b}})\citenamefont {Boseggia}, \citenamefont {Walker},
  \citenamefont {Vale}, \citenamefont {Springell}, \citenamefont {Feng},
  \citenamefont {Perry}, \citenamefont {Sala}, \citenamefont {R{\o}nnow},
  \citenamefont {Collins},\ and\ \citenamefont {McMorrow}}]{Boseggia13}%
  \BibitemOpen
  \bibfield  {author} {\bibinfo {author} {\bibfnamefont {S.}~\bibnamefont
  {Boseggia}}, \bibinfo {author} {\bibfnamefont {H.~C.}\ \bibnamefont
  {Walker}}, \bibinfo {author} {\bibfnamefont {J.}~\bibnamefont {Vale}},
  \bibinfo {author} {\bibfnamefont {R.}~\bibnamefont {Springell}}, \bibinfo
  {author} {\bibfnamefont {Z.}~\bibnamefont {Feng}}, \bibinfo {author}
  {\bibfnamefont {R.~S.}\ \bibnamefont {Perry}}, \bibinfo {author}
  {\bibfnamefont {M.~Moretti}\ \bibnamefont {Sala}}, \bibinfo {author}
  {\bibfnamefont {H.~M.}\ \bibnamefont {R{\o}nnow}}, \bibinfo {author}
  {\bibfnamefont {S.~P.}\ \bibnamefont {Collins}}, \ and\ \bibinfo {author}
  {\bibfnamefont {D.~F.}\ \bibnamefont {McMorrow}},\ }\bibfield  {title}
  {\enquote {\bibinfo {title} {Locking of iridium magnetic moments to the
  correlated rotation of oxygen octahedra in {Sr$_2$IrO$_4$} revealed by x-ray
  resonant scattering},}\ }\href@noop {} {\bibfield  {journal} {\bibinfo
  {journal} {J. Phys.: Condens. Matter}\ }\textbf {\bibinfo {volume} {25}},\
  \bibinfo {pages} {422202} (\bibinfo {year} {2013}{\natexlab{b}})}\BibitemShut
  {NoStop}%
\bibitem [{\citenamefont {Ye}\ \emph {et~al.}(2013)\citenamefont {Ye},
  \citenamefont {Chi}, \citenamefont {Chakoumakos}, \citenamefont
  {Fernandez-Baca}, \citenamefont {Qi},\ and\ \citenamefont {Cao}}]{Ye13}%
  \BibitemOpen
  \bibfield  {author} {\bibinfo {author} {\bibfnamefont {F.}~\bibnamefont
  {Ye}}, \bibinfo {author} {\bibfnamefont {S.}~\bibnamefont {Chi}}, \bibinfo
  {author} {\bibfnamefont {B.~C.}\ \bibnamefont {Chakoumakos}}, \bibinfo
  {author} {\bibfnamefont {J.~A.}\ \bibnamefont {Fernandez-Baca}}, \bibinfo
  {author} {\bibfnamefont {T.}~\bibnamefont {Qi}}, \ and\ \bibinfo {author}
  {\bibfnamefont {G.}~\bibnamefont {Cao}},\ }\bibfield  {title} {\enquote
  {\bibinfo {title} {Magnetic and crystal structures of {Sr$_2$IrO$_4$}: A
  neutron diffraction study},}\ }\href@noop {} {\bibfield  {journal} {\bibinfo
  {journal} {Phys. Rev. B.}\ }\textbf {\bibinfo {volume} {87}},\ \bibinfo
  {pages} {140406(R)} (\bibinfo {year} {2013})}\BibitemShut {NoStop}%
\bibitem [{\citenamefont {Katukuri}\ \emph
  {et~al.}(2014{\natexlab{b}})\citenamefont {Katukuri}, \citenamefont
  {Yushankhai}, \citenamefont {Siurakshina}, \citenamefont {van~den Brink},
  \citenamefont {Hozoi},\ and\ \citenamefont {Rousochatzakis}}]{Ba214_vmk_14}%
  \BibitemOpen
  \bibfield  {author} {\bibinfo {author} {\bibfnamefont {V.~M.}\ \bibnamefont
  {Katukuri}}, \bibinfo {author} {\bibfnamefont {V.}~\bibnamefont
  {Yushankhai}}, \bibinfo {author} {\bibfnamefont {L.}~\bibnamefont
  {Siurakshina}}, \bibinfo {author} {\bibfnamefont {J.}~\bibnamefont {van~den
  Brink}}, \bibinfo {author} {\bibfnamefont {L.}~\bibnamefont {Hozoi}}, \ and\
  \bibinfo {author} {\bibfnamefont {I.}~\bibnamefont {Rousochatzakis}},\
  }\bibfield  {title} {\enquote {\bibinfo {title} {Mechanism of basal-plane
  antiferromagnetism in the spin-orbit driven iridate {B}a$_2${I}r{O}$_4$},}\
  }\href@noop {} {\bibfield  {journal} {\bibinfo  {journal} {Phys. Rev. X}\
  }\textbf {\bibinfo {volume} {4}},\ \bibinfo {pages} {021051} (\bibinfo {year}
  {2014}{\natexlab{b}})}\BibitemShut {NoStop}%
\bibitem [{\citenamefont {Cao}\ \emph {et~al.}(1998)\citenamefont {Cao},
  \citenamefont {Bolivar}, \citenamefont {McCall}, \citenamefont {Crow},\ and\
  \citenamefont {Guertin}}]{Sr214_cao98}%
  \BibitemOpen
  \bibfield  {author} {\bibinfo {author} {\bibfnamefont {G.}~\bibnamefont
  {Cao}}, \bibinfo {author} {\bibfnamefont {J.}~\bibnamefont {Bolivar}},
  \bibinfo {author} {\bibfnamefont {S.}~\bibnamefont {McCall}}, \bibinfo
  {author} {\bibfnamefont {J.~E.}\ \bibnamefont {Crow}}, \ and\ \bibinfo
  {author} {\bibfnamefont {R.~P.}\ \bibnamefont {Guertin}},\ }\bibfield
  {title} {\enquote {\bibinfo {title} {Weak ferromagnetism, metal-to-nonmetal
  transition, and negative differential resistivity in single-crystal
  {Sr$_{2}$IrO$_{4}$}},}\ }\href@noop {} {\bibfield  {journal} {\bibinfo
  {journal} {Phys. Rev. B}\ }\textbf {\bibinfo {volume} {57}},\ \bibinfo
  {pages} {11039(R)} (\bibinfo {year} {1998})}\BibitemShut {NoStop}%
\bibitem [{\citenamefont {Kim}\ \emph {et~al.}(2006)\citenamefont {Kim},
  \citenamefont {Yu}, \citenamefont {Koh}, \citenamefont {Nagai}, \citenamefont
  {Ikeda}, \citenamefont {Oh},\ and\ \citenamefont {Kim}}]{Sr2RhO4_kim_2006}%
  \BibitemOpen
  \bibfield  {author} {\bibinfo {author} {\bibfnamefont {B.~J.}\ \bibnamefont
  {Kim}}, \bibinfo {author} {\bibfnamefont {J.}~\bibnamefont {Yu}}, \bibinfo
  {author} {\bibfnamefont {H.}~\bibnamefont {Koh}}, \bibinfo {author}
  {\bibfnamefont {I.}~\bibnamefont {Nagai}}, \bibinfo {author} {\bibfnamefont
  {S.~I.}\ \bibnamefont {Ikeda}}, \bibinfo {author} {\bibfnamefont {S.-J.}\
  \bibnamefont {Oh}}, \ and\ \bibinfo {author} {\bibfnamefont {C.}~\bibnamefont
  {Kim}},\ }\bibfield  {title} {\enquote {\bibinfo {title} {Missing $xy$-band
  fermi surface in 4$d$ transition-metal oxide {Sr$_2$RhO$_4$}: Effect of the
  octahedra rotation on the electronic structure},}\ }\href@noop {} {\bibfield
  {journal} {\bibinfo  {journal} {Phys. Rev. Lett.}\ }\textbf {\bibinfo
  {volume} {97}},\ \bibinfo {pages} {106401} (\bibinfo {year}
  {2006})}\BibitemShut {NoStop}%
\bibitem [{\citenamefont {Bogdanov}\ \emph {et~al.}(2013)\citenamefont
  {Bogdanov}, \citenamefont {Maurice}, \citenamefont {Rousochatzakis},
  \citenamefont {van~den Brink},\ and\ \citenamefont
  {Hozoi}}]{Os227_bogdanov_12}%
  \BibitemOpen
  \bibfield  {author} {\bibinfo {author} {\bibfnamefont {N.~A.}\ \bibnamefont
  {Bogdanov}}, \bibinfo {author} {\bibfnamefont {R.}~\bibnamefont {Maurice}},
  \bibinfo {author} {\bibfnamefont {I.}~\bibnamefont {Rousochatzakis}},
  \bibinfo {author} {\bibfnamefont {J.}~\bibnamefont {van~den Brink}}, \ and\
  \bibinfo {author} {\bibfnamefont {L.}~\bibnamefont {Hozoi}},\ }\bibfield
  {title} {\enquote {\bibinfo {title} {Magnetic state of pyrochlore
  {Cd$_2$Os$_2$O$_7$} emerging from strong competition of ligand distortions
  and longer-range crystalline anisotropy},}\ }\href@noop {} {\bibfield
  {journal} {\bibinfo  {journal} {Phys. Rev. Lett.}\ }\textbf {\bibinfo
  {volume} {110}},\ \bibinfo {pages} {127206} (\bibinfo {year}
  {2013})}\BibitemShut {NoStop}%
\bibitem [{\citenamefont {Fang}\ \emph {et~al.}(2004)\citenamefont {Fang},
  \citenamefont {Nagaosa},\ and\ \citenamefont
  {Terakura}}]{Ca2RuO4_terakura_04}%
  \BibitemOpen
  \bibfield  {author} {\bibinfo {author} {\bibfnamefont {Z.}~\bibnamefont
  {Fang}}, \bibinfo {author} {\bibfnamefont {N.}~\bibnamefont {Nagaosa}}, \
  and\ \bibinfo {author} {\bibfnamefont {K.}~\bibnamefont {Terakura}},\
  }\bibfield  {title} {\enquote {\bibinfo {title} {Orbital-dependent phase
  control in {Ca$_{2-x}$Sr$_{x}$RuO$_{4}$} $(0 < x < 0.5)$},}\ }\href@noop {}
  {\bibfield  {journal} {\bibinfo  {journal} {Phys. Rev. B}\ }\textbf {\bibinfo
  {volume} {69}},\ \bibinfo {pages} {045116} (\bibinfo {year}
  {2004})}\BibitemShut {NoStop}%
\bibitem [{\citenamefont {Lupascu}\ \emph {et~al.}(2014)\citenamefont
  {Lupascu}, \citenamefont {Clancy}, \citenamefont {Gretarsson}, \citenamefont
  {Nie}, \citenamefont {Nichols}, \citenamefont {Terzic}, \citenamefont {Cao},
  \citenamefont {Seo}, \citenamefont {Islam}, \citenamefont {Upton},
  \citenamefont {Kim}, \citenamefont {Said}, \citenamefont {Casa},
  \citenamefont {Gog}, \citenamefont {Katukuri}, \citenamefont {Stoll},
  \citenamefont {Hozoi}, \citenamefont {van~den Brink},\ and\ \citenamefont
  {Kim}}]{Ir214_lupascu_14}%
  \BibitemOpen
  \bibfield  {author} {\bibinfo {author} {\bibfnamefont {A}~\bibnamefont
  {Lupascu}}, \bibinfo {author} {\bibfnamefont {J~P}\ \bibnamefont {Clancy}},
  \bibinfo {author} {\bibfnamefont {H.}~\bibnamefont {Gretarsson}}, \bibinfo
  {author} {\bibfnamefont {Z.}~\bibnamefont {Nie}}, \bibinfo {author}
  {\bibfnamefont {J.}~\bibnamefont {Nichols}}, \bibinfo {author} {\bibfnamefont
  {J.}~\bibnamefont {Terzic}}, \bibinfo {author} {\bibfnamefont
  {G.}~\bibnamefont {Cao}}, \bibinfo {author} {\bibfnamefont {S.~S.~A.}\
  \bibnamefont {Seo}}, \bibinfo {author} {\bibfnamefont {Z.}~\bibnamefont
  {Islam}}, \bibinfo {author} {\bibfnamefont {M.~H.}\ \bibnamefont {Upton}},
  \bibinfo {author} {\bibfnamefont {J.}~\bibnamefont {Kim}}, \bibinfo {author}
  {\bibfnamefont {A.~H.}\ \bibnamefont {Said}}, \bibinfo {author}
  {\bibfnamefont {D.}~\bibnamefont {Casa}}, \bibinfo {author} {\bibfnamefont
  {T.}~\bibnamefont {Gog}}, \bibinfo {author} {\bibfnamefont {V.~M.}\
  \bibnamefont {Katukuri}}, \bibinfo {author} {\bibfnamefont {H.}~\bibnamefont
  {Stoll}}, \bibinfo {author} {\bibfnamefont {L.}~\bibnamefont {Hozoi}},
  \bibinfo {author} {\bibfnamefont {J.}~\bibnamefont {van~den Brink}}, \ and\
  \bibinfo {author} {\bibfnamefont {Y.-J.}\ \bibnamefont {Kim}},\ }\bibfield
  {title} {\enquote {\bibinfo {title} {Tuning magnetic coupling in
  {Sr$_2$IrO$_4$} thin films with epitaxial strain},}\ }\href@noop {}
  {\bibfield  {journal} {\bibinfo  {journal} {Phys. Rev. Lett.}\ }\textbf
  {\bibinfo {volume} {112}},\ \bibinfo {pages} {147201} (\bibinfo {year}
  {2014})}\BibitemShut {NoStop}%
\bibitem [{\citenamefont {Hozoi}\ \emph {et~al.}(2011)\citenamefont {Hozoi},
  \citenamefont {Siurakshina}, \citenamefont {Fulde},\ and\ \citenamefont
  {van~den Brink}}]{CuO2_dd_hozoi_11}%
  \BibitemOpen
  \bibfield  {author} {\bibinfo {author} {\bibfnamefont {L.}~\bibnamefont
  {Hozoi}}, \bibinfo {author} {\bibfnamefont {L.}~\bibnamefont {Siurakshina}},
  \bibinfo {author} {\bibfnamefont {P.}~\bibnamefont {Fulde}}, \ and\ \bibinfo
  {author} {\bibfnamefont {J.}~\bibnamefont {van~den Brink}},\ }\bibfield
  {title} {\enquote {\bibinfo {title} {{\it Ab initio} determination of {Cu}
  3$d$ orbital energies in layered copper oxides},}\ }\href@noop {} {\bibfield
  {journal} {\bibinfo  {journal} {Sci. Rep.}\ }\textbf {\bibinfo {volume}
  {1}},\ \bibinfo {pages} {65} (\bibinfo {year} {2011})}\BibitemShut {NoStop}%
\bibitem [{\citenamefont {Wu}\ \emph {et~al.}(2011)\citenamefont {Wu},
  \citenamefont {Chang}, \citenamefont {Schumann}, \citenamefont {Hu},
  \citenamefont {Cezar}, \citenamefont {Burnus}, \citenamefont {Hollmann},
  \citenamefont {Brookes}, \citenamefont {Tanaka}, \citenamefont {Braden},
  \citenamefont {Tjeng},\ and\ \citenamefont {Khomskii}}]{LaSrMnO4_wu_11}%
  \BibitemOpen
  \bibfield  {author} {\bibinfo {author} {\bibfnamefont {H.}~\bibnamefont
  {Wu}}, \bibinfo {author} {\bibfnamefont {C.~F.}\ \bibnamefont {Chang}},
  \bibinfo {author} {\bibfnamefont {O.}~\bibnamefont {Schumann}}, \bibinfo
  {author} {\bibfnamefont {Z.}~\bibnamefont {Hu}}, \bibinfo {author}
  {\bibfnamefont {J.~C.}\ \bibnamefont {Cezar}}, \bibinfo {author}
  {\bibfnamefont {T.}~\bibnamefont {Burnus}}, \bibinfo {author} {\bibfnamefont
  {N.}~\bibnamefont {Hollmann}}, \bibinfo {author} {\bibfnamefont {N.~B.}\
  \bibnamefont {Brookes}}, \bibinfo {author} {\bibfnamefont {A.}~\bibnamefont
  {Tanaka}}, \bibinfo {author} {\bibfnamefont {M.}~\bibnamefont {Braden}},
  \bibinfo {author} {\bibfnamefont {L.~H.}\ \bibnamefont {Tjeng}}, \ and\
  \bibinfo {author} {\bibfnamefont {D.~I.}\ \bibnamefont {Khomskii}},\
  }\bibfield  {title} {\enquote {\bibinfo {title} {Orbital order in
  {La$_{0.5}$Sr$_{1.5}$MnO$_4$}: Beyond a common local {Jahn}-{Teller}
  picture},}\ }\href@noop {} {\bibfield  {journal} {\bibinfo  {journal} {Phys.
  Rev. B}\ }\textbf {\bibinfo {volume} {84}},\ \bibinfo {pages} {155126}
  (\bibinfo {year} {2011})}\BibitemShut {NoStop}%
\bibitem [{\citenamefont {Braden}\ \emph {et~al.}(1998)\citenamefont {Braden},
  \citenamefont {Andr\'e}, \citenamefont {Nakatsuji},\ and\ \citenamefont
  {Maeno}}]{Ca2RuO4_braden_1998}%
  \BibitemOpen
  \bibfield  {author} {\bibinfo {author} {\bibfnamefont {M.}~\bibnamefont
  {Braden}}, \bibinfo {author} {\bibfnamefont {G.}~\bibnamefont {Andr\'e}},
  \bibinfo {author} {\bibfnamefont {S.}~\bibnamefont {Nakatsuji}}, \ and\
  \bibinfo {author} {\bibfnamefont {Y.}~\bibnamefont {Maeno}},\ }\bibfield
  {title} {\enquote {\bibinfo {title} {Crystal and magnetic structure of
  {Ca$_2$RuO$_4$}: Magnetoelastic coupling and the metal-insulator
  transition},}\ }\href@noop {} {\bibfield  {journal} {\bibinfo  {journal}
  {Phys. Rev. B}\ }\textbf {\bibinfo {volume} {58}},\ \bibinfo {pages} {847}
  (\bibinfo {year} {1998})}\BibitemShut {NoStop}%
\bibitem [{\citenamefont {Fujiyama}\ \emph {et~al.}(2014)\citenamefont
  {Fujiyama}, \citenamefont {Ohsumi}, \citenamefont {Ohashi}, \citenamefont
  {Hirai}, \citenamefont {Kim}, \citenamefont {Arima}, \citenamefont {Takata},\
  and\ \citenamefont {Takagi}}]{Sr2IrO4_NRMXD_fujiyama13}%
  \BibitemOpen
  \bibfield  {author} {\bibinfo {author} {\bibfnamefont {S.}~\bibnamefont
  {Fujiyama}}, \bibinfo {author} {\bibfnamefont {H.}~\bibnamefont {Ohsumi}},
  \bibinfo {author} {\bibfnamefont {K.}~\bibnamefont {Ohashi}}, \bibinfo
  {author} {\bibfnamefont {D.}~\bibnamefont {Hirai}}, \bibinfo {author}
  {\bibfnamefont {B.~J.}\ \bibnamefont {Kim}}, \bibinfo {author} {\bibfnamefont
  {T.}~\bibnamefont {Arima}}, \bibinfo {author} {\bibfnamefont
  {M.}~\bibnamefont {Takata}}, \ and\ \bibinfo {author} {\bibfnamefont
  {H.}~\bibnamefont {Takagi}},\ }\bibfield  {title} {\enquote {\bibinfo {title}
  {Spin and orbital contributions to magnetically ordered moments in 5$d$
  layered perovskite {Sr$_2$IrO$_4$}},}\ }\href@noop {} {\bibfield  {journal}
  {\bibinfo  {journal} {Phys. Rev. Lett.}\ }\textbf {\bibinfo {volume} {112}},\
  \bibinfo {pages} {016405} (\bibinfo {year} {2014})}\BibitemShut {NoStop}%
\bibitem [{\citenamefont {Bolvin}(2007)}]{g_bolvin_07}%
  \BibitemOpen
  \bibfield  {author} {\bibinfo {author} {\bibfnamefont {H.}~\bibnamefont
  {Bolvin}},\ }\bibfield  {title} {\enquote {\bibinfo {title} {Theoretical
  determination of the excited states and of $g$-factors of the
  {Creutz}-{Taube} ion, {[(NH$_3$)$_5$-Ru-pyrazine-Ru-(NH$_3$)$_5$]$^{5+}$}},}\
  }\href@noop {} {\bibfield  {journal} {\bibinfo  {journal} {Inorg. Chem.}\
  }\textbf {\bibinfo {volume} {46}},\ \bibinfo {pages} {417} (\bibinfo {year}
  {2007})}\BibitemShut {NoStop}%
\bibitem [{\citenamefont {Ohtomo}\ and\ \citenamefont
  {Hwang}(2004)}]{polar_layers_ohtomo_04}%
  \BibitemOpen
  \bibfield  {author} {\bibinfo {author} {\bibfnamefont {A}~\bibnamefont
  {Ohtomo}}\ and\ \bibinfo {author} {\bibfnamefont {H~Y}\ \bibnamefont
  {Hwang}},\ }\bibfield  {title} {\enquote {\bibinfo {title} {A high-mobility
  electron gas at the {LaAlO$_3$}/{SrTiO$_3$} heterointerface},}\ }\href@noop
  {} {\bibfield  {journal} {\bibinfo  {journal} {Nature}\ }\textbf {\bibinfo
  {volume} {427}},\ \bibinfo {pages} {423} (\bibinfo {year}
  {2004})}\BibitemShut {NoStop}%
\bibitem [{\citenamefont {Koster}\ \emph {et~al.}(2008)\citenamefont {Koster},
  \citenamefont {Brinkman}, \citenamefont {Hilgenkamp}, \citenamefont
  {Rijnders},\ and\ \citenamefont {Blank}}]{polar_layers_koster_08}%
  \BibitemOpen
  \bibfield  {author} {\bibinfo {author} {\bibfnamefont {G.}~\bibnamefont
  {Koster}}, \bibinfo {author} {\bibfnamefont {A.}~\bibnamefont {Brinkman}},
  \bibinfo {author} {\bibfnamefont {H.}~\bibnamefont {Hilgenkamp}}, \bibinfo
  {author} {\bibfnamefont {A.~J. H.~M.}\ \bibnamefont {Rijnders}}, \ and\
  \bibinfo {author} {\bibfnamefont {D.~H.~A.}\ \bibnamefont {Blank}},\
  }\bibfield  {title} {\enquote {\bibinfo {title} {High- {T$_c$}
  superconducting thin films with composition control on a sub-unit cell level;
  the effect of the polar nature of the cuprates},}\ }\href@noop {} {\bibfield
  {journal} {\bibinfo  {journal} {J. Phys.: Condens. Matter}\ }\textbf
  {\bibinfo {volume} {20}},\ \bibinfo {pages} {264007} (\bibinfo {year}
  {2008})}\BibitemShut {NoStop}%
\bibitem [{\citenamefont {Ning}\ \emph {et~al.}(2013)\citenamefont {Ning},
  \citenamefont {Wang},\ and\ \citenamefont {Zhang}}]{interf_RuNi_ning13}%
  \BibitemOpen
  \bibfield  {author} {\bibinfo {author} {\bibfnamefont {Xingkun}\ \bibnamefont
  {Ning}}, \bibinfo {author} {\bibfnamefont {Zhanjie}\ \bibnamefont {Wang}}, \
  and\ \bibinfo {author} {\bibfnamefont {Zhidong}\ \bibnamefont {Zhang}},\
  }\bibfield  {title} {\enquote {\bibinfo {title} {{Exchange bias effect and
  large coercivity enhancement in {S}rRuO$_3$/NiO multilayers}},}\ }\href@noop
  {} {\bibfield  {journal} {\bibinfo  {journal} {J. Phys. D: Appl. Phys.}\
  }\textbf {\bibinfo {volume} {46}},\ \bibinfo {pages} {452001} (\bibinfo
  {year} {2013})}\BibitemShut {NoStop}%
\bibitem [{\citenamefont {de~Graaf}\ \emph {et~al.}(1999)\citenamefont
  {de~Graaf}, \citenamefont {Sousa},\ and\ \citenamefont
  {Broer}}]{qc_NNs_degraaf_99}%
  \BibitemOpen
  \bibfield  {author} {\bibinfo {author} {\bibfnamefont {C.}~\bibnamefont
  {de~Graaf}}, \bibinfo {author} {\bibfnamefont {C.}~\bibnamefont {Sousa}}, \
  and\ \bibinfo {author} {\bibfnamefont {R.}~\bibnamefont {Broer}},\ }\bibfield
   {title} {\enquote {\bibinfo {title} {Ionization and excitation energies in
  {CuCl} and {NiO} within different embedding schemes},}\ }\href@noop {}
  {\bibfield  {journal} {\bibinfo  {journal} {J. Mol. Struct. (Theochem)}\
  }\textbf {\bibinfo {volume} {458}},\ \bibinfo {pages} {53} (\bibinfo {year}
  {1999})}\BibitemShut {NoStop}%
\bibitem [{\citenamefont {Hozoi}\ \emph {et~al.}(2002)\citenamefont {Hozoi},
  \citenamefont {de~Vries}, \citenamefont {van Oosten}, \citenamefont {Broer},
  \citenamefont {Cabrero},\ and\ \citenamefont {de~Graaf}}]{Na2V2O5_hozoi_02}%
  \BibitemOpen
  \bibfield  {author} {\bibinfo {author} {\bibfnamefont {L.}~\bibnamefont
  {Hozoi}}, \bibinfo {author} {\bibfnamefont {A.~H.}\ \bibnamefont {de~Vries}},
  \bibinfo {author} {\bibfnamefont {A.~B.}\ \bibnamefont {van Oosten}},
  \bibinfo {author} {\bibfnamefont {R.}~\bibnamefont {Broer}}, \bibinfo
  {author} {\bibfnamefont {J.}~\bibnamefont {Cabrero}}, \ and\ \bibinfo
  {author} {\bibfnamefont {C.}~\bibnamefont {de~Graaf}},\ }\bibfield  {title}
  {\enquote {\bibinfo {title} {Theoretical characterization of the ground and
  optically excited states of $\alpha'$-{NaV$_2$O$_5$}},}\ }\href@noop {}
  {\bibfield  {journal} {\bibinfo  {journal} {Phys. Rev. Lett.}\ }\textbf
  {\bibinfo {volume} {89}},\ \bibinfo {pages} {076407} (\bibinfo {year}
  {2002})}\BibitemShut {NoStop}%
\bibitem [{\citenamefont {Domingo}\ \emph {et~al.}(2012)\citenamefont
  {Domingo}, \citenamefont {Rodr\'{i}guez-Fortea}, \citenamefont {Swart},
  \citenamefont {de~Graaf},\ and\ \citenamefont {Broer}}]{NiO_domingo_12}%
  \BibitemOpen
  \bibfield  {author} {\bibinfo {author} {\bibfnamefont {A.}~\bibnamefont
  {Domingo}}, \bibinfo {author} {\bibfnamefont {A.}~\bibnamefont
  {Rodr\'{i}guez-Fortea}}, \bibinfo {author} {\bibfnamefont {M.}~\bibnamefont
  {Swart}}, \bibinfo {author} {\bibfnamefont {C.}~\bibnamefont {de~Graaf}}, \
  and\ \bibinfo {author} {\bibfnamefont {R.}~\bibnamefont {Broer}},\ }\bibfield
   {title} {\enquote {\bibinfo {title} {{\it Ab initio} absorption spectrum of
  {NiO} combining molecular dynamics with the embedded cluster approach in a
  discrete reaction field},}\ }\href@noop {} {\bibfield  {journal} {\bibinfo
  {journal} {Phys. Rev. B}\ }\textbf {\bibinfo {volume} {85}},\ \bibinfo
  {pages} {155143} (\bibinfo {year} {2012})}\BibitemShut {NoStop}%
\bibitem [{\citenamefont {Maurice}\ \emph {et~al.}(2013)\citenamefont
  {Maurice}, \citenamefont {Verma}, \citenamefont {Zadrozny}, \citenamefont
  {Luo}, \citenamefont {Borycz}, \citenamefont {Long}, \citenamefont
  {Truhlar},\ and\ \citenamefont {Gagliardi}}]{SIA_Fe_maurice_2013}%
  \BibitemOpen
  \bibfield  {author} {\bibinfo {author} {\bibfnamefont {R.}~\bibnamefont
  {Maurice}}, \bibinfo {author} {\bibfnamefont {P.}~\bibnamefont {Verma}},
  \bibinfo {author} {\bibfnamefont {J.~M.}\ \bibnamefont {Zadrozny}}, \bibinfo
  {author} {\bibfnamefont {S.}~\bibnamefont {Luo}}, \bibinfo {author}
  {\bibfnamefont {J.}~\bibnamefont {Borycz}}, \bibinfo {author} {\bibfnamefont
  {J.~R.}\ \bibnamefont {Long}}, \bibinfo {author} {\bibfnamefont {Donald~G.}\
  \bibnamefont {Truhlar}}, \ and\ \bibinfo {author} {\bibfnamefont
  {L.}~\bibnamefont {Gagliardi}},\ }\bibfield  {title} {\enquote {\bibinfo
  {title} {Single-ion magnetic anisotropy and isotropic magnetic couplings in
  the metal-organic framework {Fe$_2$(dobdc)}},}\ }\href@noop {} {\bibfield
  {journal} {\bibinfo  {journal} {Inorg. Chem.}\ }\textbf {\bibinfo {volume}
  {52}},\ \bibinfo {pages} {9379} (\bibinfo {year} {2013})}\BibitemShut
  {NoStop}%
\bibitem [{\citenamefont {Figgen}\ \emph {et~al.}(2009)\citenamefont {Figgen},
  \citenamefont {Peterson}, \citenamefont {Dolg},\ and\ \citenamefont
  {Stoll}}]{ECP_stoll_5d_09}%
  \BibitemOpen
  \bibfield  {author} {\bibinfo {author} {\bibfnamefont {D.}~\bibnamefont
  {Figgen}}, \bibinfo {author} {\bibfnamefont {K.~A.}\ \bibnamefont
  {Peterson}}, \bibinfo {author} {\bibfnamefont {M.}~\bibnamefont {Dolg}}, \
  and\ \bibinfo {author} {\bibfnamefont {H.}~\bibnamefont {Stoll}},\ }\bibfield
   {title} {\enquote {\bibinfo {title} {Energy-consistent pseudopotentials and
  correlation consistent basis sets for the 5$d$ elements {Hf}--{Pt}},}\
  }\href@noop {} {\bibfield  {journal} {\bibinfo  {journal} {J. Chem. Phys.}\
  }\textbf {\bibinfo {volume} {130}},\ \bibinfo {eid} {164108} (\bibinfo {year}
  {2009})}\BibitemShut {NoStop}%
\bibitem [{\citenamefont {Dunning}(1989)}]{BS_dunning_89}%
  \BibitemOpen
  \bibfield  {author} {\bibinfo {author} {\bibfnamefont {T.~H.}\ \bibnamefont
  {Dunning}},\ }\bibfield  {title} {\enquote {\bibinfo {title} {Gaussian basis
  sets for use in correlated molecular calculations. {I.} the atoms boron
  through neon and hydrogen},}\ }\href@noop {} {\bibfield  {journal} {\bibinfo
  {journal} {J. Chem. Phys.}\ }\textbf {\bibinfo {volume} {90}},\ \bibinfo
  {pages} {1007} (\bibinfo {year} {1989})}\BibitemShut {NoStop}%
\bibitem [{\citenamefont {Pierloot}\ \emph {et~al.}(1995)\citenamefont
  {Pierloot}, \citenamefont {Dumez}, \citenamefont {Widmark},\ and\
  \citenamefont {Roos}}]{BS_ano_pierloot_95}%
  \BibitemOpen
  \bibfield  {author} {\bibinfo {author} {\bibfnamefont {K.}~\bibnamefont
  {Pierloot}}, \bibinfo {author} {\bibfnamefont {B.}~\bibnamefont {Dumez}},
  \bibinfo {author} {\bibfnamefont {P.-O.}\ \bibnamefont {Widmark}}, \ and\
  \bibinfo {author} {\bibfnamefont {B.}~\bibnamefont {Roos}},\ }\bibfield
  {title} {\enquote {\bibinfo {title} {Density matrix averaged atomic natural
  orbital ({ANO}) basis sets for correlated molecular wave functions},}\
  }\href@noop {} {\bibfield  {journal} {\bibinfo  {journal} {Theor. Chim.
  Acta}\ }\textbf {\bibinfo {volume} {90}},\ \bibinfo {pages} {87} (\bibinfo
  {year} {1995})}\BibitemShut {NoStop}%
\bibitem [{\citenamefont {Fuentealba}\ \emph {et~al.}(1985)\citenamefont
  {Fuentealba}, \citenamefont {von Szentpaly}, \citenamefont {Preuss},\ and\
  \citenamefont {Stoll}}]{ECP_stoll_alkaline_85}%
  \BibitemOpen
  \bibfield  {author} {\bibinfo {author} {\bibfnamefont {P.}~\bibnamefont
  {Fuentealba}}, \bibinfo {author} {\bibfnamefont {L.}~\bibnamefont {von
  Szentpaly}}, \bibinfo {author} {\bibfnamefont {H.}~\bibnamefont {Preuss}}, \
  and\ \bibinfo {author} {\bibfnamefont {H.}~\bibnamefont {Stoll}},\ }\bibfield
   {title} {\enquote {\bibinfo {title} {Pseudopotential calculations for
  alkaline-earth atoms},}\ }\href@noop {} {\bibfield  {journal} {\bibinfo
  {journal} {J. Phys. B}\ }\textbf {\bibinfo {volume} {18}},\ \bibinfo {pages}
  {1287} (\bibinfo {year} {1985})}\BibitemShut {NoStop}%
\bibitem [{\citenamefont {Werner}\ and\ \citenamefont
  {Knowles}(1988)}]{CI_molpro_jcp_88}%
  \BibitemOpen
  \bibfield  {author} {\bibinfo {author} {\bibfnamefont {H.-J.}\ \bibnamefont
  {Werner}}\ and\ \bibinfo {author} {\bibfnamefont {P.~J.}\ \bibnamefont
  {Knowles}},\ }\bibfield  {title} {\enquote {\bibinfo {title} {An efficient
  internally contracted multiconfiguration-reference configuration interaction
  method},}\ }\href@noop {} {\bibfield  {journal} {\bibinfo  {journal} {J.
  Chem. Phys.}\ }\textbf {\bibinfo {volume} {89}},\ \bibinfo {pages} {5803}
  (\bibinfo {year} {1988})}\BibitemShut {NoStop}%
\bibitem [{\citenamefont {Knowles}\ and\ \citenamefont
  {Werner}(1992)}]{CI_molpro_tca_92}%
  \BibitemOpen
  \bibfield  {author} {\bibinfo {author} {\bibfnamefont {P.~J.}\ \bibnamefont
  {Knowles}}\ and\ \bibinfo {author} {\bibfnamefont {H.-J.}\ \bibnamefont
  {Werner}},\ }\bibfield  {title} {\enquote {\bibinfo {title} {Internally
  contracted multiconfiguration-reference configuration interaction
  calculations for excited states},}\ }\href@noop {} {\bibfield  {journal}
  {\bibinfo  {journal} {Theor. Chim. Acta}\ }\textbf {\bibinfo {volume} {84}},\
  \bibinfo {pages} {95} (\bibinfo {year} {1992})}\BibitemShut {NoStop}%
\bibitem [{\citenamefont {Pipek}\ and\ \citenamefont
  {Mezey}(1989)}]{localization_PM}%
  \BibitemOpen
  \bibfield  {author} {\bibinfo {author} {\bibfnamefont {J.}~\bibnamefont
  {Pipek}}\ and\ \bibinfo {author} {\bibfnamefont {P.~G.}\ \bibnamefont
  {Mezey}},\ }\bibfield  {title} {\enquote {\bibinfo {title} {A fast intrinsic
  localization procedure applicable for {\it ab initio} and semiempirical
  linear combination of atomic orbital wave functions},}\ }\href@noop {}
  {\bibfield  {journal} {\bibinfo  {journal} {J. Chem. Phys.}\ }\textbf
  {\bibinfo {volume} {90}},\ \bibinfo {pages} {4916} (\bibinfo {year}
  {1989})}\BibitemShut {NoStop}%
\bibitem [{\citenamefont {Neese}(2012)}]{orca}%
  \BibitemOpen
  \bibfield  {author} {\bibinfo {author} {\bibfnamefont {Frank}\ \bibnamefont
  {Neese}},\ }\bibfield  {title} {\enquote {\bibinfo {title} {The {ORCA}
  program system},}\ }\href {\doibase 10.1002/wcms.81} {\bibfield  {journal}
  {\bibinfo  {journal} {Wiley Interdiscip. Rev. Comput. Mol. Sci.}\ }\textbf
  {\bibinfo {volume} {2}},\ \bibinfo {pages} {73--78} (\bibinfo {year}
  {2012})}\BibitemShut {NoStop}%
\bibitem [{\citenamefont {Chibotaru}\ and\ \citenamefont
  {Ungur}(2012)}]{g_chibotaru_12}%
  \BibitemOpen
  \bibfield  {author} {\bibinfo {author} {\bibfnamefont {L.~F.}\ \bibnamefont
  {Chibotaru}}\ and\ \bibinfo {author} {\bibfnamefont {L.}~\bibnamefont
  {Ungur}},\ }\bibfield  {title} {\enquote {\bibinfo {title} {{\it Ab initio}
  calculation of anisotropic magnetic properties of complexes. {I.} unique
  definition of pseudospin hamiltonians and their derivation},}\ }\href@noop {}
  {\bibfield  {journal} {\bibinfo  {journal} {J. Chem. Phys.}\ }\textbf
  {\bibinfo {volume} {137}},\ \bibinfo {pages} {064112} (\bibinfo {year}
  {2012})}\BibitemShut {NoStop}%
\bibitem [{\citenamefont {Gerloch}\ and\ \citenamefont
  {McMeeking}(1975)}]{gerloch_mcmeeking_DaltonTrans_1975}%
  \BibitemOpen
  \bibfield  {author} {\bibinfo {author} {\bibfnamefont {M.}~\bibnamefont
  {Gerloch}}\ and\ \bibinfo {author} {\bibfnamefont {R.~F.}\ \bibnamefont
  {McMeeking}},\ }\bibfield  {title} {\enquote {\bibinfo {title} {Paramagnetic
  properties of unsymmetrical transition-metal complexes},}\ }\href {\doibase
  10.1039/DT9750002443} {\bibfield  {journal} {\bibinfo  {journal} {J. Chem.
  Soc.{,} Dalton Trans.}\ ,\ \bibinfo {pages} {2443}} (\bibinfo {year}
  {1975})}\BibitemShut {NoStop}%
\bibitem [{\citenamefont {Pantazis}\ \emph {et~al.}(2008)\citenamefont
  {Pantazis}, \citenamefont {Chen}, \citenamefont {Landis},\ and\ \citenamefont
  {Neese}}]{Orca_Irbasis}%
  \BibitemOpen
  \bibfield  {author} {\bibinfo {author} {\bibfnamefont {D.~A.}\ \bibnamefont
  {Pantazis}}, \bibinfo {author} {\bibfnamefont {X.~Y.}\ \bibnamefont {Chen}},
  \bibinfo {author} {\bibfnamefont {C.~R.}\ \bibnamefont {Landis}}, \ and\
  \bibinfo {author} {\bibfnamefont {F.}~\bibnamefont {Neese}},\ }\bibfield
  {title} {\enquote {\bibinfo {title} {All-electron scalar relativistic basis
  sets for third-row transition metal atoms},}\ }\href@noop {} {\bibfield
  {journal} {\bibinfo  {journal} {J. Chem. Theory Comput.}\ }\textbf {\bibinfo
  {volume} {4}},\ \bibinfo {pages} {908} (\bibinfo {year} {2008})}\BibitemShut
  {NoStop}%
\bibitem [{\citenamefont {Angeli}\ \emph {et~al.}(2001)\citenamefont {Angeli},
  \citenamefont {Cimiraglia},\ and\ \citenamefont {Malrieu}}]{NEVPT2_CPL_2001}%
  \BibitemOpen
  \bibfield  {author} {\bibinfo {author} {\bibfnamefont {C.}~\bibnamefont
  {Angeli}}, \bibinfo {author} {\bibfnamefont {R.}~\bibnamefont {Cimiraglia}},
  \ and\ \bibinfo {author} {\bibfnamefont {J.-P.}\ \bibnamefont {Malrieu}},\
  }\bibfield  {title} {\enquote {\bibinfo {title} {N-electron valence state
  perturbation theory: a fast implementation of the strongly contracted
  variant},}\ }\href {\doibase http://dx.doi.org/10.1016/S0009-2614(01)01303-3}
  {\bibfield  {journal} {\bibinfo  {journal} {Chem. Phys. Lett.}\ }\textbf
  {\bibinfo {volume} {350}},\ \bibinfo {pages} {297} (\bibinfo {year}
  {2001})}\BibitemShut {NoStop}%
\bibitem [{\citenamefont {Schapiro}\ \emph {et~al.}(2013)\citenamefont
  {Schapiro}, \citenamefont {Sivalingam},\ and\ \citenamefont
  {Neese}}]{NEVPT2_benchmark_2013}%
  \BibitemOpen
  \bibfield  {author} {\bibinfo {author} {\bibfnamefont {I.}~\bibnamefont
  {Schapiro}}, \bibinfo {author} {\bibfnamefont {K.}~\bibnamefont
  {Sivalingam}}, \ and\ \bibinfo {author} {\bibfnamefont {F.}~\bibnamefont
  {Neese}},\ }\bibfield  {title} {\enquote {\bibinfo {title} {Assessment of
  n-electron valence state perturbation theory for vertical excitation
  energies},}\ }\href {\doibase 10.1021/ct400136y} {\bibfield  {journal}
  {\bibinfo  {journal} {J. Chem. Theory Comput.}\ }\textbf {\bibinfo {volume}
  {9}},\ \bibinfo {pages} {3567} (\bibinfo {year} {2013})}\BibitemShut
  {NoStop}%
\bibitem [{\citenamefont {Katukuri}\ \emph
  {et~al.}(2014{\natexlab{c}})\citenamefont {Katukuri}, \citenamefont
  {Nishimoto}, \citenamefont {Yushankhai}, \citenamefont {Stoyanova},
  \citenamefont {Kandpal}, \citenamefont {Choi}, \citenamefont {Coldea},
  \citenamefont {Rousochatzakis}, \citenamefont {Hozoi},\ and\ \citenamefont
  {van~den Brink}}]{Ir213_katukuri_14}%
  \BibitemOpen
  \bibfield  {author} {\bibinfo {author} {\bibfnamefont {V.~M.}\ \bibnamefont
  {Katukuri}}, \bibinfo {author} {\bibfnamefont {S.}~\bibnamefont {Nishimoto}},
  \bibinfo {author} {\bibfnamefont {V.}~\bibnamefont {Yushankhai}}, \bibinfo
  {author} {\bibfnamefont {A.}~\bibnamefont {Stoyanova}}, \bibinfo {author}
  {\bibfnamefont {H.}~\bibnamefont {Kandpal}}, \bibinfo {author} {\bibfnamefont
  {S.~K.}\ \bibnamefont {Choi}}, \bibinfo {author} {\bibfnamefont
  {R.}~\bibnamefont {Coldea}}, \bibinfo {author} {\bibfnamefont
  {I.}~\bibnamefont {Rousochatzakis}}, \bibinfo {author} {\bibfnamefont
  {L.}~\bibnamefont {Hozoi}}, \ and\ \bibinfo {author} {\bibfnamefont
  {J.}~\bibnamefont {van~den Brink}},\ }\bibfield  {title} {\enquote {\bibinfo
  {title} {{K}itaev interactions between $j = 1/2$ moments in honeycomb
  {Na$_2$IrO$_3$} are large and ferromagnetic: insights from {\it ab} {\it
  initio} quantum chemistry calculations},}\ }\href@noop {} {\bibfield
  {journal} {\bibinfo  {journal} {New J. Phys.}\ }\textbf {\bibinfo {volume}
  {16}},\ \bibinfo {pages} {013056} (\bibinfo {year}
  {2014}{\natexlab{c}})}\BibitemShut {NoStop}%
\end{thebibliography}

%

\end{document}